\documentclass[floatfix,superscriptaddress,longbibliography,nofootinbib]{revtex4-2}
\RequirePackage[sort&compress]{natbib}
\usepackage{color}
\usepackage{float}
\usepackage[table]{xcolor}
\usepackage{amssymb,amsfonts,amsmath}
\usepackage{bm}
\usepackage[pdftex]{graphicx}
\usepackage{xcolor}
\usepackage{comment}
\usepackage{multirow}
\usepackage{booktabs}

\usepackage[colorlinks = true,
            linkcolor = blue,
            urlcolor  = blue,
            citecolor = blue,
            anchorcolor = blue]{hyperref}

\usepackage{dcolumn}
\usepackage{bm}
\usepackage{xcolor}

\usepackage{xcolor}
\definecolor{RoyalBlue}{HTML}{4169e1}
\definecolor{ForestGreen}{HTML}{228b22}

\usepackage{setspace}
\usepackage{comment}

\usepackage{todonotes}

\begin{document}

\title{Mixing Individual and Collective Behaviours to Predict Out-of-Routine Mobility}

\author{Sebastiano Bontorin}
\affiliation{Fondazione Bruno Kessler, Via Sommarive 18, 38123 Povo (TN), Italy}
\affiliation{Department of Physics, University of Trento, Via Sommarive 14, 38123 Povo (TN), Italy}

\author{Simone Centellegher}
\affiliation{Fondazione Bruno Kessler, Via Sommarive 18, 38123 Povo (TN), Italy}

\author{Riccardo Gallotti}
\affiliation{Fondazione Bruno Kessler, Via Sommarive 18, 38123 Povo (TN), Italy}

\author{Luca Pappalardo}
\affiliation{ISTI - National Research Council, Via Giuseppe Moruzzi 1, 56127 Pisa (PI), Italy}

\author{Bruno Lepri}
\affiliation{Fondazione Bruno Kessler, Via Sommarive 18, 38123 Povo (TN), Italy}

\author{Massimiliano Luca}
\email[Corresponding author:~]{mluca@fbk.eu}%
\affiliation{Fondazione Bruno Kessler, Via Sommarive 18, 38123 Povo (TN), Italy}

\date{\today}

\begin{abstract}
Predicting human displacements is crucial for addressing various societal challenges, including urban design, traffic congestion, epidemic management, and migration dynamics. 
While predictive models like deep learning and Markov models offer insights into individual mobility, they often struggle with out-of-routine behaviours. 
Our study introduces an approach that dynamically integrates individual and collective mobility behaviours, leveraging collective intelligence to enhance prediction accuracy. Evaluating the model on millions of privacy-preserving trajectories across three US cities, we demonstrate its superior performance in predicting out-of-routine mobility, surpassing even advanced deep learning methods. 
Spatial analysis highlights the model's effectiveness near urban areas with a high density of points of interest, where collective behaviours strongly influence mobility. 
During disruptive events like the COVID-19 pandemic, our model retains predictive capabilities, unlike individual-based models. 
By bridging the gap between individual and collective behaviours, our approach offers transparent and accurate predictions, crucial for addressing contemporary mobility challenges.
\end{abstract}

\maketitle

\section*{Significance Statement}
Crowds significantly influence individual decisions, as evidenced by studies on collective intelligence, social psychology, and behavioural economics. Existing models that predict individual patterns struggle to predict unexpected behaviours as they tend to memorise individual preferences and lack generalisation capabilities. Our approach mixes collective and individual decisions to overcome these limitations while offering interpretability and stronger performance in out-of-routine mobility prediction. We find that collective information is crucial in forecasting these unconventional choices by capturing broader patterns and trends that individual-based models overlook. In addition, we find that collective information makes the model robust to external shocks and reveals that out-of-routine mobility patterns close to areas with a higher density of points of interest are more predictable and more localised.

\section{Introduction}

Understanding human mobility patterns is relevant to many pressing problems in our societies \cite{pappalardo2023future}, including the design of sustainable and livable cities \cite{lee2015relating, bohm2022gross}, traffic congestion avoidance \cite{akhtar2021review, wang2012understanding}, epidemics spread mitigation and public health monitoring \cite{wesolowski2012quantifying, xiong2020mobile, kraemer2020effect, lucchini2021living}, urban and socioeconomic segregation \cite{yabe2023behavioral, moro2021mobility}, and migration management after natural disasters, economic shocks, and wars \cite{lu2012predictability, deville2014dynamic, gonzalez2008understanding, bosetti2020heterogeneity, klamser2023}.

The task of predicting individuals' future whereabouts, often referred to as next location prediction \cite{luca2021survey, chekol2022survey}, has attracted particular interest in light of the growing availability of extensive mobility data and the development of advanced statistical techniques \cite{barbosa2018human, luca2021survey, pappalardo2023future}.
On the one hand, sophisticated deep learning solutions have gained substantial attention given their capacity to uncover complex patterns from extensive datasets \cite{de2015artificial,liu2016predicting,feng2018deepmove,xue2021mobtcast,luca2021survey}.
However, these models often lack interpretability, functioning as black boxes that obscure the underlying mechanisms driving predictions \cite{pappalardo2023future, luca2021survey}.
On the other hand, simple and interpretable models such as Markov models allow the analysis of the mechanisms behind the predictions but often exhibit lower accuracy in forecasting future movements \cite{ashbrook2002learning,gambs2012next,calabrese2010human, zhao2021characteristics}. 

Both deep learning and Markov models are trained on individual mobility trajectories, which are typically inherently predictable as people tend to visit previously visited locations at regular times \cite{song2010limits,de2013interdependence,lu2013approaching,pappalardo2015returners,cuttone2018understanding,smolak2021impact,barbosa2018human}.
However, in some cases, individuals may have a marked preference for exploring new destinations \cite{pappalardo2015returners,cuttone2018understanding,scherrer2018travelers,amichi2020understanding, schlapfer2021universal} or be forced to alter their routine due to external factors such as job loss, health issues, natural disasters, or epidemics \cite{toole2015tracking,almaatouq2016mobile,barbosa2021uncovering,centellegher2024longterm, haug2020ranking,zhang2021impact, wilson2016rapid,hong2021measuring,gray2012natural, lucchini2021living}.
Predicting such out-of-routine mobility is a challenge for statistical models because, being designed to capture regular patterns in individual trajectories, they often memorise training data rather than learning generalised mobility behaviours \cite{luca2021survey, luca2023trajectory}.
A large body of literature on human behaviour across various contexts, such as social networks \cite{mason2012collaborative, jayles2017social, becker2017network, almaatouq2020adaptive}, financial networks \cite{kaustia2012peer, pan2012decoding, heimer2016peer}, and voting and political polarization \cite{coleman2004effect, becker2019wisdom, shi2019wisdom}, indicates that an individual's decisions are significantly influenced by the behaviour of the group or community they are exposed to \cite{surowiecki2005wisdom, dyer2009leadership, woolley2010evidence}.
This suggests that information about collective behaviours also holds predictive power for individuals.
However, the potential of combining individual and collective behaviours to enhance human mobility prediction remains largely unexplored.

In this paper, we bridge this gap by introducing a new approach for next location prediction that dynamically integrates individual and collective mobility behaviours, leveraging the expected predictability of an individual's movement. 
We evaluate our model's performance using the trajectories of millions of anonymized, opted-in individuals across three US cities over an eight-month period and compare its accuracy with approaches relying only on individual or collective data.
By offering transparency without compromising predictive accuracy, our model demonstrates considerable generalisation capabilities, particularly in out-of-routine mobility behaviours, surpassing more sophisticated deep learning methods.
Notably, a detailed spatial analysis reveals that our model particularly benefits from collective information in out-of-routine mobility prediction, especially in areas with a high density of points of interest. 
While individual-based models are significantly affected by disruptions to recurrent mobility patterns, our model's dynamic integration of individual and collective behaviours enables it to maintain predictive capabilities during disruptive events such as the COVID-19 pandemic.

\section*{Results}
\begin{figure*}[h!]
\centering
\includegraphics[width=\linewidth]{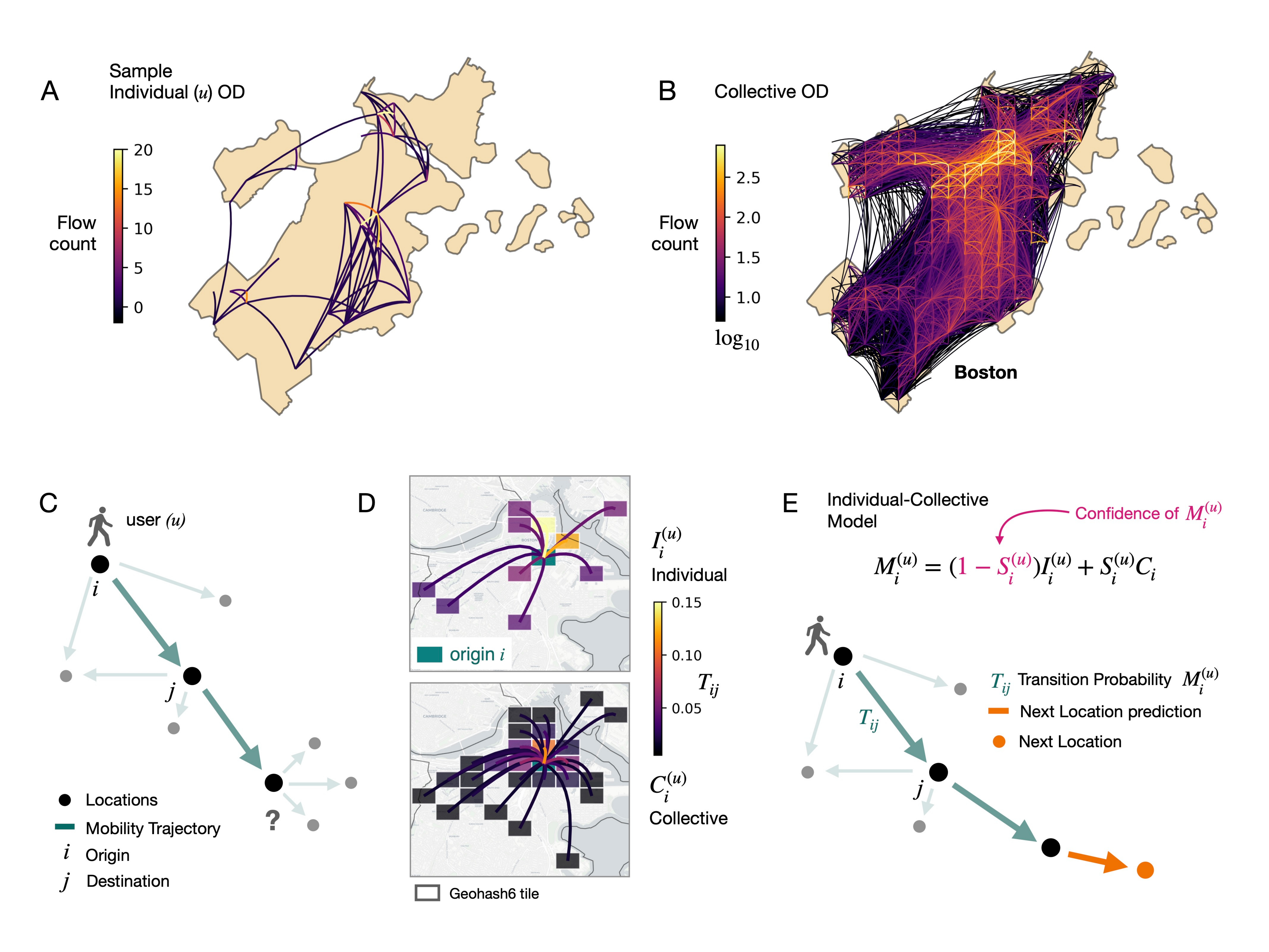}
\caption{\textbf{Dynamic interplay of individual and collective mobility.} \textbf{(A)} An individual origin-destination matrix for a synthetic individual $u$. \textbf{(B)} The collective origin-destination (OD) matrix computed for Boston using GPS trajectories.
\textbf{(C)} An individual trajectory for a user $u$ starting from location $i$. 
Next location prediction consists of predicting $u$'s next visited location. 
\textbf{(D)} The set of $u$'s historical trajectories (panel A) is used to define the transition probabilities $I^{(u)}_{i}$ from location $i$.
$C_i$ represents the probability distribution of all transitions made by any user starting from location $i$, generated from the OD matrix (panel B). Destinations' locations $j$ are coloured based on their visitation probability, $T_{ij}$, from origin $i$.
\textbf{(E)} $M_i^{(u)}$'s prediction of individual $u$'s next location is performed by dynamically combining $I^{(u)}_{i}$ and $C_i$, based on the normalised Shannon entropy $S^{(u)}_i$ computed from the mobility trajectories of $u$. 
Maps: Stamen Maps. Icons: Fontawesome.
}
\label{fig:Fig0}
\end{figure*}
A spatio-temporal point is a pair $p = (i, t)$, where $i$ represents a geographic location and $t$ the time of the visit. 
We define a trajectory $P = \{ p_1,p_2,\dots,p_n \}$ as a daily time-ordered sequence of $n$ spatio-temporal points. 
Each individual user $u$ has a set of $N$ historical trajectories $\mathcal{H}^{(u)} = \{P_1, \dots, P_N\}$ from which we compute $I^{(u)}_i$, representing the set of transition probabilities of user $u$ starting from location $i$ (see Fig.~\ref{fig:Fig0}A and Methods for details). 
By aggregating the trajectories of all individuals, we calculate the collective origin-destination matrix $C$ (see Fig.~\ref{fig:Fig0}B), with $C_i$ representing the probability distribution of all transitions made by any individual starting from location $i$. 

Given an individual's trajectory $P \in \mathcal{H}^{(u)}$, next location prediction is the problem of forecasting the next point $p_{n+1} \in P$ \cite{luca2021survey, luca2023trajectory, chekol2022survey}. 
$I^{(u)}_i$ and $C_i$ are Markov-based solutions to next location predictors \cite{gambs2012next, chekol2022survey}.

We define a model $M_i^{(u)}$ that dynamically combines  $I^{(u)}_i$ and $C_i$ based on the predictability of $u$'s next location from origin $i$.
When the next location is highly predictable based on $\mathcal{H}^{(u)}$, the model relies more on individual information in $I^{(u)}_i$ for the prediction. 
Conversely, the model relies more on the collective information in $C_i$ when the next location is hard to predict.
To quantify the predictability of $u$'s next location from $i$, we employ the normalised Shannon's entropy of $I^{(u)}_i$ \cite{song2010limits, eagle2009eigenbehaviors, pappalardo2016analytical}:
\begin{equation}
    S^{(u)}_i = - \frac{\sum_{k \in L^{(u)}} I^{(u)}_{i,k} \cdot \log(I^{(u)}_{i,k})}{\log |L^{(u)}|} 
\label{eq:shannon}
\end{equation}
\noindent where $L^{(u)}$ is the set of distinct locations visited by $u$, $\log |L^{(u)}|$ is a normalisation factor so that $S_i^{(u)} \in [0, 1]$, and $I^{(u)}_{i,k}$ is the probability of $u$ moving from location $i$ to location $k$.
$S_i^{(u)}$ is high when the locations $u$ visited from $i$ have similar visitation probabilities, indicating a diverse range of destinations;  $S_i^{(u)}$ is low when $u$ predominantly visits one location from $i$, indicating a marked individual preference for a specific destination.
We use $S_i^{(u)}$ to combine probabilities from $I^{(u)}_i$ and $C_i$ as follows:
\begin{equation}
    M_i^{(u)} = (1 - S^{(u)}_i) I_i^{(u)} + S^{(u)}_i C_i
    \label{eq:ic-od}
\end{equation}

\noindent where $1 - S^{(u)}_i$ is the confidence of model $M_{i}^{(u)}$ in relying on individual information.
To derive Markov transition probabilities, we normalise $M_i^{(u)}, \forall i$ using the softmax function:
$$\text{softmax}(l)_i = \frac{e^{l_i}}{\sum_{j=1}^n e^{l_j}}$$
where $l_1,l_2, \dots l_n$ are the transition probabilities of $M_i^{(u)}$ in Equation \ref{eq:ic-od}. 
When a transition from location $i$ is not represented in $\mathcal{H}^{(u)}$, the probability distribution $I^{(u)}_i$ is empty, and we set $S^{(u)}_i = 1$ to indicate maximum uncertainty.
In such instances, no historical data is available for location $i$, and the model prediction relies solely on collective information in $C_i$.
Fig.~\ref{fig:Fig0}C-E illustrates how model $M_i^{(u)}$ works.

\begin{figure}[!]
\centering
\includegraphics[width=\linewidth]{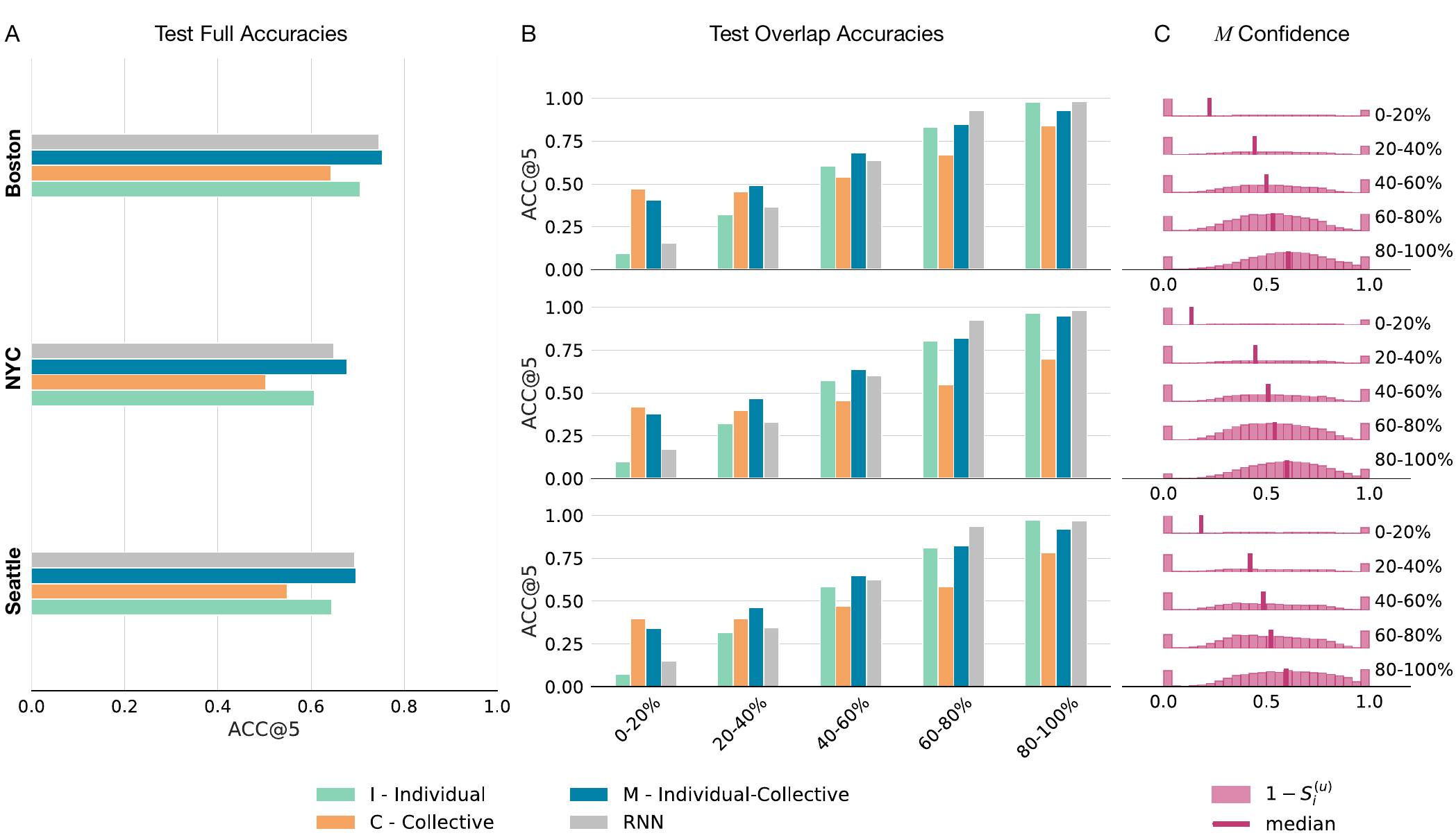}
\caption{\textbf{Accuracy of the models.} 
Top-5 accuracy (ACC@5) for Boston, New York City (NYC) and Seattle using models $I$, $C$, and $M$.
\textbf{(A)} ACC@5 on the full test set.  $M$ shows a performance better than $I$ and $C$ and comparable to Recurrent Neural Networks (RNNs).
\textbf{(B)} Models are tested against different train-test overlap scenarios, with 0-20\% describing out-of-routine mobility and 80-100\% routinary mobility behaviour. 
$M$ shows improvements in accuracy over $I$ in smaller overlaps, where test trajectories mostly consist of novel transitions never observed during training.
\textbf{(C)} Distributions of $M$'s confidence, represented by $1 - S^{(u)}_i$, in relying on individual information $I$. 
In the case of out-of-routine behaviours (low overlaps), the lower median value of $1 - S^{(u)}_i$ indicates less reliance of $M$ on individual information $I$. In this scenario, collective behaviours $C$ enhance the predictive capabilities of $M$.
The peaks observed around $1-S^{(u)}_{i} = 0$ result from instances of transitions from a location $i$ that is not represented in the training trajectories of user $u$. 
In such a case, we set $S^{(u)}_i = 1$, forcing $M$ to rely only on $C$.}
\label{fig:accuracies}
\end{figure}

\begin{table*}
\centering
\resizebox{\textwidth}{!}{
\begin{tabular}{@{}lrrr @{\hspace*{5mm}}rrr @{\hspace*{5mm}}rrr @{\hspace*{5mm}}rrr @{\hspace*{5mm}}rrr @{\hspace*{5mm}}rrr@{}}
\toprule
& \multicolumn{3}{c}{\textbf{full test set}} & \multicolumn{3}{c}{\textbf{0-20}} & \multicolumn{3}{c}{\textbf{20-40}} & \multicolumn{3}{c}{\textbf{40-60}} & \multicolumn{3}{c}{\textbf{60-80}} & \multicolumn{3}{c}{\textbf{80-100}} \\
\midrule
& \textbf{$I$} & \textbf{$C$} & \textbf{$M$} & \textbf{$I$} & \textbf{$C$} & \textbf{$M$} & \textbf{$I$} & \textbf{$C$} & \textbf{$M$} & \textbf{$I$} & \textbf{$C$} & \textbf{$M$} & \textbf{$I$} & \textbf{$C$} & \textbf{$M$} & \textbf{$I$} & \textbf{$C$} & \textbf{$M$} \\
\cmidrule(l){2-19}

 \textbf{NYC} & 0.6083 & 0.5035 & \bf 0.6784 & 0.096 & \bf 0.416 & 0.376 & 0.319 & 0.398 & \bf 0.468 & 0.572 & 0.453 & \bf  0.637 & 0.801 & 0.546 & \bf 0.817 & \bf  0.966 & 0.698 & 0.948 \\
 \textbf{Boston} & 0.7062 & 0.6436 & \bf 0.7536 & 0.093 & \bf 0.468 & 0.407 & 0.32 & 0.454 &\bf  0.492 & 0.604 & 0.537 &\bf  0.68 & 0.831 & 0.669 & \bf 0.847 &\bf  0.977 & 0.839 & 0.929 \\
 \textbf{Seattle} & 0.6453 & 0.5496 & \bf  0.6972 & 0.073 & \bf 0.395 & 0.34 & 0.314 & 0.397 &\bf  0.461 & 0.581 & 0.471 &\bf  0.646 & 0.808 & 0.584 & \bf 0.824 &\bf  0.971 & 0.781 & 0.918 \\
 
\bottomrule
\end{tabular}%
}
\caption{\textbf{Accuracy of models.} Performance of models $I$, $C$, $M$ on all trajectories in the full test set and the trajectories stratified based on their overlap with training trajectories using the Longest Common Sub-Trajectory (LCST). 
}
\label{tab:ic-accuracies}
\end{table*}

We employ the notations $I$, $C$, and $M$ to represent the models exploiting individual, collective, and their combination.
To derive $I$ and $C$ for real individuals, we use privacy-enhanced GPS trajectories collected in Boston, Seattle, and New York City (NYC) from January 3rd to March 1st, 2020 (see Supplementary Note S1, Supplementary Table S1, Supplementary Figure S1).
We tessellate the cities into rectangular tiles (locations) of $1.2$ km $\times$ $609.4$ m, corresponding to GeoHashes of level 6. 
These tiles are the basis for mapping user's stops and computing users' trajectories (see Methods).
To address potential under- and over-representation issues, we filter out trajectories with less than four points ($|P^{(u)}| < 4$), users with fewer than two trajectories ($|\mathcal{H}^{(u)}| < 2$), and remove the top 95th percentile of the most represented users (see Supplementary Note S1, Supplementary Figure S3).
Fig.~\ref{fig:Fig0}A-B shows examples of $I$ and $C$ derived from the GPS dataset for the city of Boston.

For each individual $u$, we allocate 80\% of their least recent trajectories for model training, while the 20\% most recent trajectories form the test set. 
During the training phase, for each location $i$ and user $u$, we compute $I^{(u)}_i$, $C_i$, and $S^{(u)}_i$. 
Subsequently, we perform next location predictions on the test set and assess the models' performance using the top-5 accuracy metric (ACC@5). 
ACC@5 is a standard metric for evaluating next location prediction tasks and represents the percentage of instances where the correct next location is among the top five predicted destinations \cite{luca2023trajectory, luca2021survey}.

Table \ref{tab:ic-accuracies} and Fig.~\ref{fig:accuracies}A show that $M$ significantly enhances accuracy compared to $I$ and $C$ and obtains comparable performances with deep learning baselines like Recurrent Neural Networks (RNNs) \cite{goodfellow2016deep, luca2021survey}.
$M$ exhibits relative improvements in ACC@5 over $I$ of +15\% in NYC, +13\% in Seattle, and +12\% in Boston. 
Furthermore, $M$ outperforms $C$ with improvements of +35\% in NYC, +31\% in Seattle, and +21\% in Boston.
Despite relying on a large number of parameters, RNNs exhibit only marginal relative improvements over $M$: +2\% in NYC, +2\% in Seattle, and +1\% in Boston (see Supplementary Note S6, Supplementary Table S3). 
Therefore, the dynamic interplay between individual and collective information enables model $M$ to significantly enhance the overall predictive performance compared to models relying exclusively on individual ($I$) or collective ($C$) information.

\subsection*{Models' generalisation capability}

The recurrent patterns in human mobility often result in significant similarity among individual trajectories \cite{barbosa2018human, schneider2013unravelling, luca2023trajectory}. Consequently, our test set may include a significant portion of trajectories already present in the training set. 
Recent research emphasises the importance of considering trajectory overlap between the training and test sets for a comprehensive evaluation of next location predictors, as it significantly influences the assessment of a model's generalisation capability \cite{luca2023trajectory}.

We quantify trajectory train-test overlap with the Longest Common Sub-Trajectory (LCST) \cite{luca2023trajectory}, which evaluates shared sub-sequences between two trajectories, considering the order and frequency of visits.
LCST ranges between 0 and 1 and measures how much of a test trajectory the model has already observed in the training set. 
A high LCST indicates the presence of recurrent mobility patterns in an individual's trajectories, while
a low LCST indicates instances of an individual's out-of-routine mobility behaviours.
We stratify test trajectories into bins based on LCST, spanning overlap ranges of 0-20\%, 20-40\%, 40-60\%, 60-80\%, and 80-100\% (see Methods for details).
For example, a test trajectory falls within the 0-20\% bin when its maximum overlap with any training trajectory has an LCST $\in [0,0.2]$.
The distribution of test trajectories across the bins is not uniform, as bins 0-20\% and 80-100\% contain fewer trajectories than bins 20-80\% (see Supplementary Note S1, Supplementary Figure S3).

Table \ref{tab:ic-accuracies} and Fig.~\ref{fig:accuracies}B provide models' performance within each trajectory overlap bin.
$M$ significantly outperforms $I$ and $C$ for intermediate levels of overlap, with improvements across cities up to +16\% (20-40\% overlap), +13\% (40-60\% overlap), and around $+2$\% (60-80\% overlap). 
$C$ surpasses $M$ for 0-20\% overlap only, with improvements up to 16\% across the cities, while
$I$ outperforms $M$ by only around 1\% for 80-100\% overlap.
Fig.~\ref{fig:accuracies}C shows that the median value of $1 - S^{(u)}_i$ distribution increases with trajectory overlap, i.e., the higher the trajectory overlap, the higher the confidence of model $M$ on $I$ and, consequently, the lower the reliance on $C$. 
In other words, when historical information about the individual is unreliable due to significant differences from the current trajectory (low trajectory overlap), using collective information is crucial for making reasonable predictions. 
When the current trajectory closely resembles those in the training set (high trajectory overlap), it is best to rely on individual information.
Note that RNNs achieve a low accuracy on low trajectory overlap bins and high accuracy on high overlap bins (Fig.~\ref{fig:accuracies}B), confirming that deep learning models struggle to predict out-of-routine mobility.
To verify the robustness of our findings, we carry out the same analysis using more fine-grained tessellations, obtaining comparable results (see Supplementary Note S6).

\begin{figure*}[!ht]
    \centering
    \includegraphics[width=1.0\linewidth]{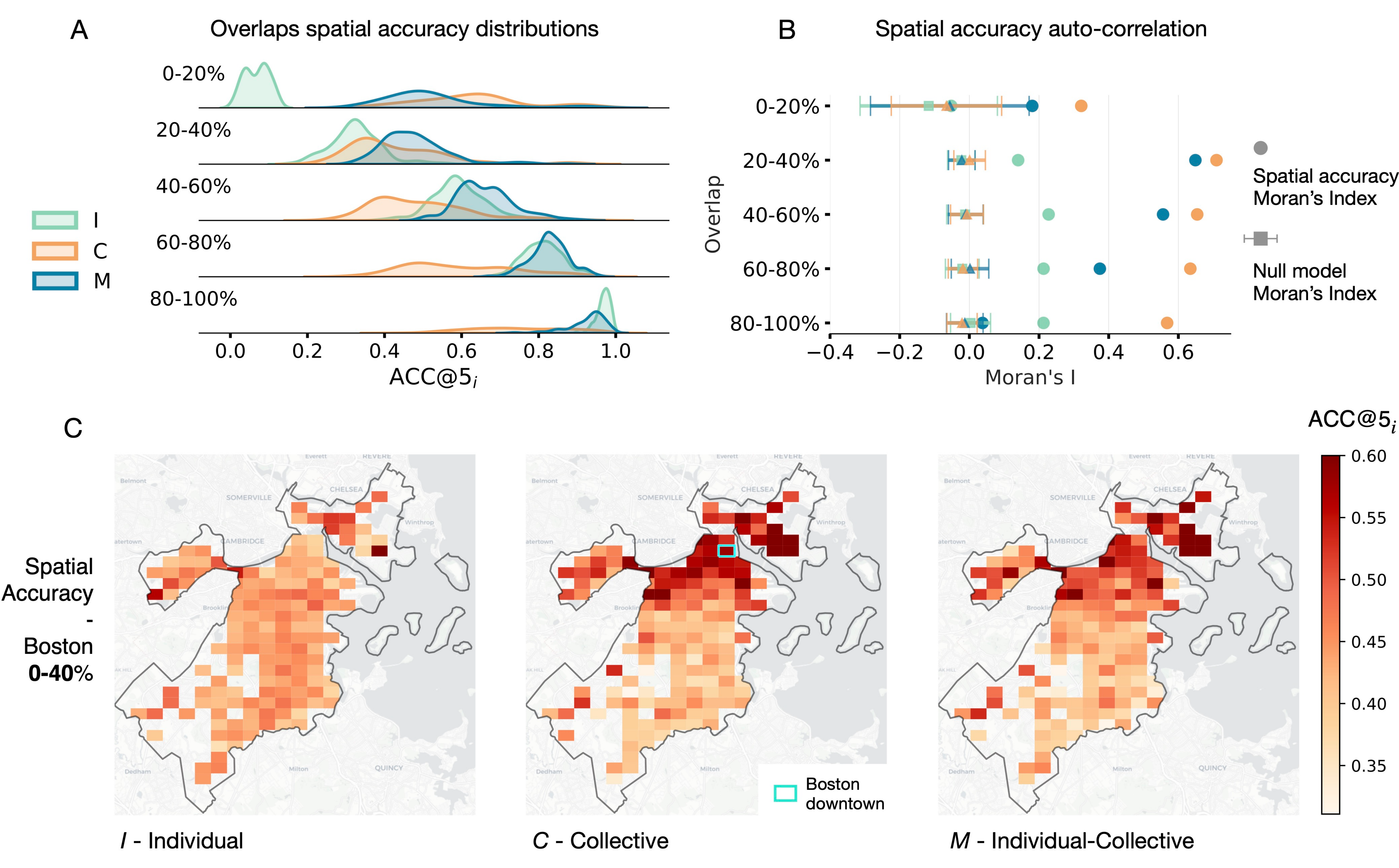}
    \caption{\textbf{Spatial distributions of accuracies.} 
    \textbf{(A)} Distribution of accuracies for $I$, $C$ and $M$ in predicting movements from a location $i$ (ACC@5$_{i}$). 
    As the test set includes more out-of-routine movements (e.g., 0-40\% overlap), model $M$ aided by collective information $C$ performs better. Conversely, the individual model $I$ provides better predictions when tested on routinary trajectories (high overlaps). \textbf{(B)} Spatial autocorrelation of the models' accuracies in corresponding overlaps quantified via the Moran's index. For low overlaps, such as 0-40\%, model $C$ exhibits clustered accuracy (large Moran's index). \textbf{(C)} Spatial distribution of ACC@5$_{i}$ in Boston for $I$, $C$ and $M$ in the 0-40\% overlap.
    Notably, for $C$ and $M$, areas with higher accuracies are concentrated in proximity to downtown (upper centre) and Boston Logan International Airport (upper right). 
    Maps: Stamen Maps.}
    \label{fig:spatial_acc}
\end{figure*}

\subsection*{Spatial properties of models' accuracy}

To investigate potential spatial dependencies in the accuracy of $I$, $C$ and $M$, we compute ACC@5 specifically for the subset of test transitions originating from a location $i$. We refer to it as ACC@5$_i$.
In Fig.~\ref{fig:spatial_acc}A, we present the spatial distributions of ACC@5$_i$ for the city of Boston. 
We find similar distributions for Seattle and NYC (see Supplementary Note S2, Supplementary Figure S4, Supplementary Figure S5).

In scenarios with low trajectory overlap (0-40\%), the ACC@5$_i$ distributions for $M$ and $C$ are consistently shifted to higher values compared to $I$ (see Fig.~\ref{fig:spatial_acc}A).
While model $C$ performs best for trajectories with 0-20\% overlap, model $M$ shows the highest performance in the 20-40\% overlap range, benefiting from collective information from $C$.
As trajectory overlap increases, the ACC@5$_i$ distributions for $M$ and $I$ become more similar, while the distribution for $C$ widens and remains skewed towards lower accuracy (see Fig.~\ref{fig:spatial_acc}A).

Fig.~\ref{fig:spatial_acc}C shows that when out-of-routine is predominant (0-40\% overlap), there is considerable spatial heterogeneity in the distribution of ACC@5$_i$ across different locations. This is particularly pronounced when using collective information for predictions, as in $C$ and $M$.
We quantify this spatial property using Moran's index \cite{anselin1995local, chen2015new} and find that models $C$ and $M$ exhibit a significant and positive spatial autocorrelation, indicating that locations where the models are accurate are spatially close (see Fig.~\ref{fig:spatial_acc}B-C and Supplementary Fig.~S4-S5).

In particular, we observe that locations with high accuracies for $C$ and $M$ are clustered in the proximity of critical urban areas.
For instance, in Boston around downtown and Logan International Airport (see Fig.~\ref{fig:spatial_acc}C), in NYC in Manhattan, and in Seattle around downtown (see Supplementary Fig. S4-S5).

\begin{figure*}[!ht]
    \centering
    \includegraphics[width=\linewidth]{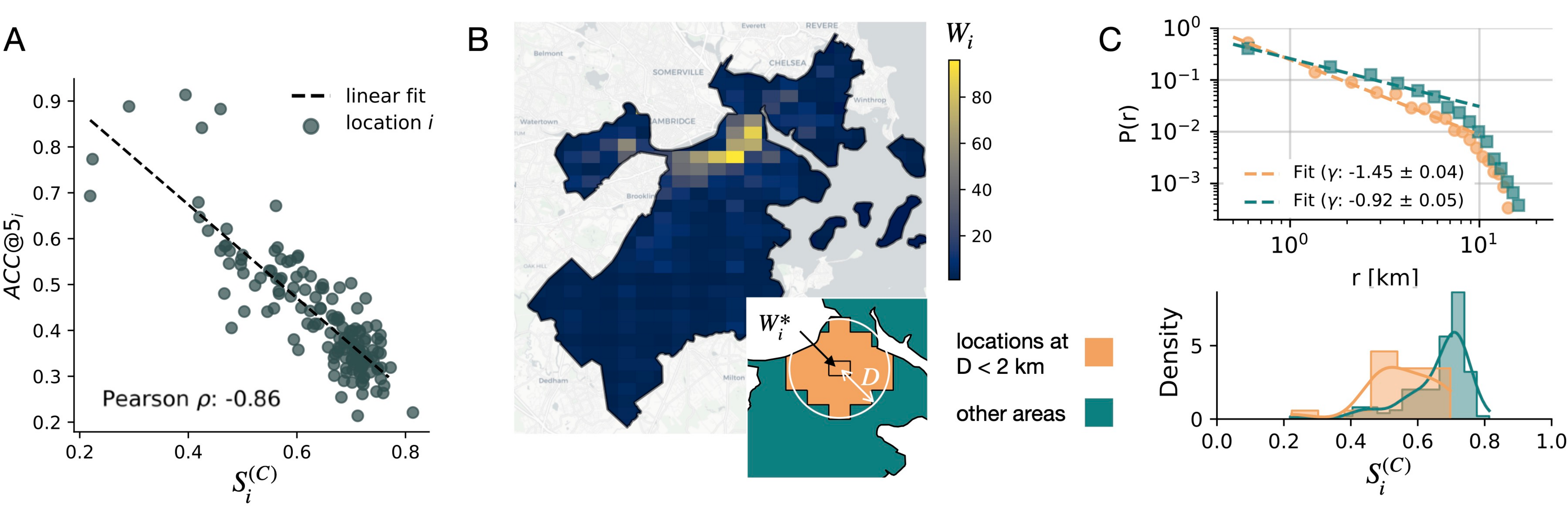}
    \caption{\textbf{Statistical properties of collective mobility in Boston.} \textbf{(A)} Accuracy of $C$ (ACC@5$_{i}$) from a location $i$ versus its entropy $S_{i}^{(C)}$. 
    We find a negative correlation (Pearson $\rho = -0.86$). 
    \textbf{(B)} Spatial distribution of the number of Points of Interest (POIs) per location, $W_{i}$ (extracted from OpenStreetMap).
    \textbf{(Inset)} In orange, locations within a distance $D = 2$ km from the location $i^*$ with the largest number of POIs ($W_{i}^*$). 
    In green, locations farther away from $i^*$ by a distance greater than $D$. 
    \textbf{(C)} Distribution of travel distances, $P(r)$, distinguishing between origins within the two aforementioned areas and fitted with a power-law function \cite{barbosa2018human} in the interval of 0 to 10 km. 
    The exponent of $\gamma = -1.45 \pm 0.04$ underscores the prevalence of localised mobility when individuals are in proximity to $i^*$, while in other areas we have an exponent of $\gamma = -0.92 \pm 0.04$. 
    This result indicates that mobility near POIs tends to be more concentrated and less spatially dispersed towards specific destinations. 
    This behaviour is further corroborated by the entropy $S_{i}^{(C)}$ distribution, which skews towards lower values and indicates mobility directed towards specific tiles. Maps: Stamen Maps.}
    \label{fig:col_properties}
\end{figure*}

The observed spatial autocorrelation for $C$ and $M$ suggests a potential relationship between collective mobility behaviours and spatially clustered urban factors when out-of-routine mobility is predominant.
To verify this hypothesis, we measure the predictability of collective behaviours from a location $i$ as the entropy of collective mobility: 
\begin{equation}
    S^{(C)}_i = - \frac{\sum_{k \in L^{(C)}} C_{ik} \cdot \log(C_{ik})}{\log |L^{(C)}|} \in [0, 1]
\end{equation}
where $L^{(C)}$ is the set of unique locations in $C$. 
$S_i^{(C)}$ is high when locations people visited from $i$ have similar visitation probabilities, indicating a diverse range of destinations;  $S_i^{(C)}$ is low when people predominantly visit one location from $i$, indicating a marked collective preference for a specific destination.
$S^{(C)}_i$ is strongly negatively correlated with ACC@5$_i$ (Pearson correlation of $\rho = -0.85$), indicating that locations from which $C$ provides the most accurate next location predictions are those with the lowest entropy $S^{(C)}_{i}$ (see Fig.~\ref{fig:col_properties}A and Supplementary Figure S8-S9).
Notably, we find that locations with low $S_{i}^{(C)}$ are clustered in proximity to specific urban areas, which we hypothesise to be locations hosting key commercial, financial, and cultural venues.

To verify this hypothesis, we first collect from OpenStreetMap \cite{OpenStreetMap} the number of points of interest (POIs) in each location $i$, $W_{i}$. 
We then split each city into two areas: one comprising locations within a geographical distance $D$ from the location $i^*$ with the maximum number of POIs ($W_{i}^*$), and the other consisting of locations farther away from $i^*$ by a distance greater than $D$ (see Fig.~\ref{fig:col_properties}B for $D = 2$ km in Boston).

We find that the distribution of $S_{i}^{(C)}$ skews towards lower values when location $i$ is within distance $D$ to $i^*$ (orange area in Fig.~\ref{fig:col_properties}B).
Additionally, individuals' movements that originate within this distance tend to travel shorter distances than those originating farther away (see Fig.~\ref{fig:col_properties}C).
These results support our hypothesis that mobility in dense areas of POIs is less spatially dispersed, with the probability of travelling to a destination concentrated in a small subset of locations. 
We see similar results in Seattle and NYC (Supplementary Figure S6-S7). 
For statistical robustness, we verify that these results are not a consequence of possible biases, such as areas close to POIs having a larger sample size of transitions in the dataset (see Supplementary Note S4 and Supplementary Figure S10-S14). 

\begin{table*}
\centering
\resizebox{\textwidth}{!}{%
\begin{tabular}{@{}lrrr @{\hspace*{5mm}}rrr @{\hspace*{5mm}}rrr @{\hspace*{5mm}}rrr @{\hspace*{5mm}}rrr@{}}
\toprule
         & \multicolumn{3}{c}{\textbf{Mar. - Apr.}} & \multicolumn{3}{c}{\textbf{Apr. - May}} & \multicolumn{3}{c}{\textbf{May - Jun.}} & \multicolumn{3}{c}{\textbf{Jun. - Jul.}} & \multicolumn{3}{c}{\textbf{Jul. - Aug.}} \\
\midrule
         & \textbf{I }        & \textbf{C}        & \textbf{M}       & \textbf{I }        & \textbf{C}        & \textbf{M}                
         & \textbf{I }        & \textbf{C}        & \textbf{M}       & \textbf{I }        & \textbf{C}        & \textbf{M}        
         & \textbf{I }        & \textbf{C}        & \textbf{M}        \\
\cmidrule(l){2-16}

\textbf{New York} & 0.484 &	0.541 & 0.655 & 0.390 &	0.541 &	0.617 & 0.358 & 0.536 & 0.602 & 0.325 & 0.524 & 0.585 & 0.290 & 0.515 & 0.570   \\
\textbf{Boston}   & 0.491 & 0.632 & 0.696 & 0.394 & 0.609 & 0.654 & 0.350 & 0.620 & 0.651 & 0.307 & 0.619 & 0.642 & 0.255 & 0.615 & 0.637 \\
\textbf{Seattle}  & 0.424 & 0.578 & 0.649 & 0.352 & 0.582 & 0.629 & 0.311 & 0.583 & 0.619 & 0.271 & 0.557 & 0.596 & 0.236 & 0.543 & 0.579  \\

\bottomrule
\end{tabular}%
}
\caption{\textbf{Models accuracy during the COVID-19 pandemics}. ACC@5 of $I$, $C$ and $M$ trained on trajectories collected before March 1st, evaluated on trajectories during and after lockdown every month after the pandemic declaration (March 11th). 
The COVID-19 pandemic acts as a disruptive event that introduces behavioural changes in human mobility. 
The individual model $I$ loses half of its predictive capabilities, while the models relying on collective behaviours ($C$ and $M$) generalise even after months, maintaining accuracies comparable to the pre-COVID-19 period.}
\label{tab:dist_shift_accs_5}
\end{table*}

\subsection*{Model reliability under COVID-19 restrictions}

The COVID-19 pandemic significantly altered people's mobility patterns \cite{zhang2021impact, lucchini2021living, haug2020ranking, flaxman2020nature, chinazzi2020effect}, with non-pharmaceutical interventions inducing a shift in how people moved and visited locations \cite{yabe2023behavioral, napoli2023socioeconomic, lucchini2021living}. In this context, we investigate the reliability of the models during these behavioural shifts. 

We train $I$, $C$, and $M$ using trajectories recorded until March 11th, 2020 and evaluate their performance on a test set consisting of five months of data collected between March 11th, 2020 (WHO's pandemic declaration \cite{world2022director}) and August 11th, 2020. 
For each month in the test set, we report models' ACC@5 in Table \ref{tab:dist_shift_accs_5}.

In each city, $I$ exhibits a notable decline in performance as the months go by, losing $44.16$\% of its predictive power by the time of testing trajectories collected between July and August. 
$C$'s performance also degrades over time but more moderately, with an average drop of $4.51$\% between the first and last month. 
$M$ exhibits a moderate average decrease of $5.32$\%, which is slightly worse than $C$ but notably better than $I$.
These results highlight that models based on individual-level information are less resilient to behavioural shifts than models based on collective information. 
As a consequence, combining individual information with collective one allows for enhanced resilience against disruptive behavioural changes.

\section*{Discussion}

Crowds influence individual decisions, a phenomenon extensively documented in studies on collective intelligence, social psychology, and behavioural economics \cite{cialdini2007influence, venema2020doubt}. 
In various scenarios, individuals may face uncertainty or possess limited knowledge, prompting them to defer to collective decisions for guidance. 
We leverage these insights in a parameter-free approach to next location prediction that dynamically integrates individual and collective information. 
Our model is designed to harness the power of collective behaviours alongside individual patterns, offering a holistic framework for predicting human whereabouts.

Our model offers a potential solution to the limitations inherent in current state-of-the-art models, including sophisticated deep learning approaches. 
Indeed, while deep learning models excel in predicting routinary movements \cite{luca2021survey}, they suffer from two primary drawbacks: opacity and limited interpretability \cite{pappalardo2023future, luca2021survey}, and difficulties in forecasting out-of-routine choices because of their tendency to memorise patterns observed in individual trajectories \cite{luca2023trajectory}.
Our approach overcomes these shortcomings by offering full interpretability and strong performance in out-of-routine movement prediction, even during disruptive events like the COVID-19 pandemic \cite{luca2023trajectory}.
Collective information is particularly crucial in anticipating these out-of-routine choices, enabling us to capture patterns and trends that may not be apparent at the individual level alone.
In the context of the ongoing discourse on mechanistic versus deep learning models for human mobility \cite{barbosa2018human, luca2021survey}, our study suggests the necessity to integrate dynamic mechanisms like ours into deep learning frameworks.
This integration is essential for achieving predictions that are both interpretable and accurate in scenarios involving both routinary and out-of-routine movements.

Our study additionally reveals that predictions relying on collective information, especially the most accurate ones, spatially cluster around urban areas characterised by a dense concentration of points of interest.
These areas are where the probability of travelling to a destination is highly concentrated within a smaller subset of locations.
These findings align with recent research on flow generation, underscoring that the presence and nature of points of interest play a significant role in shaping human mobility \cite{simini2021deep}. 
An intriguing research challenge lies in enhancing our model to incorporate an additional mechanism that accounts for the density of points of interest in the area where the individual is currently located. 

While conventional predictive models for human mobility often prioritise capturing routine movements, our approach paves the road for dynamic models that integrate information at individual and collective levels to enable robust predictions even for out-of-routine mobility behaviours.

\section*{Methods}

\subsection*{Stop Location Detection} 
For our experiments, we use a privacy-enhanced GPS dataset provided by Cuebiq as part of the Data for Good COVID-19 Collaborative program. 
This dataset encompasses privacy-enhanced GPS locations spanning nine months (January to August 2020) in New York City, Seattle, and Boston. 
The data originates from approximately two million anonymous users who willingly opted to share their information anonymously for research purposes, adhering to the guidelines of the CCPA (California Consumer Privacy Act) compliant framework. In addition to anonymizng the data, the data provider removes sensitive points of interest from the dataset, and obfuscates inferred home locations to the Census Block Group level.

We extract individual user stops from the dataset through the following procedure. 
Initially, we identify each temporal sequence of GPS coordinates within a 65-meter radius, where a user stayed for a minimum of 5 minutes \cite{hariharan2004project}. Subsequently, we apply the DBSCAN algorithm \cite{ester1996density} to identify dense clusters of points within a distance of $\epsilon = \Delta_s - 5$.
We define these dense clusters as stop locations. 
For a more detailed explanation of the GPS data processing, refer to~\cite{lucchini2021living}.

\subsection*{Points of Interest}
We download from OpenStreetMap (OSM) data about points of interest (POIs) in New York City, Seattle, and Boston.
POIs describe public venues such as restaurants, and parks in a city. 
We employ a dictionary of amenities covering a large set of public venues (see Supplementary Note S1 for the specific entries used). 
We compute the number of POIs extracted from OSM in each GeoHash tile $i$ and refer to it as $W_{i}$. The maps of POIs for Boston, Seattle and New York City can be seen in Supplementary Fig.~S6.

\subsection*{Origin-destination matrix} \label{sec:icmodel}

Each individual $u$ in our datasets is associated with a set of consecutive trajectories $\mathcal{H}^{(u)} = \{P_{1}, \dots, P_{N}\}$, each capturing the locations visited over a 24-hour period.
The origin-destination matrix of $u$ captures the transition probability, $T^{(u)}_{ij} / T^{(u)}_{i}$, between each pair of locations visited by $u$, where $T^{(u)}_{ij}$ is the total number of transitions in $\mathcal{H}^{(u)}$ from location $i$ to location $j$, and $T^{(u)}_{i} = \sum_{j \in L} T^{(u)}_{ij}$ accounts for the total transitions from location $i$ to any other location in the set $L^{(u)}$ of locations visited by $u$ \cite{schneider2013unravelling}.

\subsection*{Train-Test overlap}

To compute the overlap between two trajectories, $P^{(u)} = \{p_1, p_2, \dots, p_n\}$  and $R^{(u)} = \{r_1, r_2, \dots, r_m\}$, we introduce the prefix $P_i$ of $P^{(u)}$ as the list of the first $i$-th locations in $P^{(u)}$, i.e., $P_i = \{p_1, \dots, p_i\}$ (dropping index $^{(u)}$ for simplicity). 
The definition extends similarly to $R$. 
The size of the Longest Common Sub-Trajectory for two prefixes 
$P_i$ and $R_j$ is denoted as LCST ($f$ in Equation \ref{eq:LCST}) and is defined as follows:

\begin{equation}
    \small
    \begin{aligned}
        &f(P_i, R_j) = \\
        &\begin{cases} 
            0, &\text{if } i=0 \text{ or } j=0 \\ 
            f(P_{i-1}, R_{j-1})+1, &\text{if } i,j > 0 \text{ and } p_{i} = r_{j} \\
            \max(f(P_{i-1}, R_{j}), f(P_{i}, R_{j-1})), &\text{if } i,j > 0 \text{ and } p_{i} \neq r_{j}
        \end{cases}
    \end{aligned}
\label{eq:LCST}
\end{equation}

We quantify the overlap between a test trajectory $R$ and the training set as the maximum LCST over all the trajectories in the training set:

\begin{equation}
\max_{P \in \mathcal{H}^{(u)}} \text{LCST}(P, R)
\end{equation}

We normalise the LCST score within the range $[0, 1]$, and we assign each trajectory to one of the following five bins: 0-20\%, 20-40\%, 40-60\%, 60-80\%, and 80-100\%, based on their LCST score \cite{luca2023trajectory}.
For instance, a test trajectory with an LCST score less than or equal to $0.2$ will be assigned to the bin of trajectories with 0-20\% LCST. We exclude trajectories with an exact 0\% overlap as they are primarily consisting of individuals remaining in the same location, for a small set of locations. Refer to Supplementary Note S1, Supplementary Figure S2, Supplementary Table S2, Supplementary Figure S3 for further details.
The distribution of LCST for test trajectories is detailed in Supplementary Note S1 and Supplementary Figure S2.


\begin{acknowledgements}
The authors would like to thank Cuebiq for kindly providing us with the mobility dataset for this research through their Data for Good program. L.P. has been supported by 1) PNRR (Piano
Nazionale di Ripresa e Resilienza) in the context of the research program 20224CZ5X4\_PE6\_PRIN 2022 “URBAI
-- Urban Artificial Intelligence” (CUP B53D23012770006),
funded by European Union -- Next Generation EU; 2) EU
project H2020 SoBigData++ G.A. 871042; 3) NextGenerationEU—National Recovery and Resilience Plan (Piano
Nazionale di Ripresa e Resilienza, PNRR), Project “SoBigData.it—Strengthening the Italian RI for Social Mining
and Big Data Analytics”, prot. IR0000013, avviso n. 3264
on 28/12/2021. 
The work of S.C., R.G., B.L. and M.L. has been supported by the PNRR ICSC National Research Centre for High Performance Computing, Big Data and Quantum Computing (CN00000013), under the NRRP MUR program funded by the NextGenerationEU. B.L. also acknowledges the support of the PNRR project FAIR - Future AI Research (PE00000013), under the NRRP MUR program funded by the NextGenerationEU and by the European Union’s Horizon Europe research and innovation program under grant agreement No. 101120237 (ELIAS).
\end{acknowledgements}

\subsection*{Contributions}
M.L. and S.B. designed and developed the model. S.C. processed the data and reviewed the model. M.L., S.B., and S.C. performed the experiments. M.L. designed the study. M.L., S.B., S.C., L.P., and B.L. contributed to interpreting the results and writing the paper. L.P. coordinated the writing of the paper. R.G. read and approved the paper. 

\subsection*{Competing Financial Interests}
The authors declare no competing financial interests

\subsection*{Data Availability}
The data used in this work are publicly available from the original references

\subsection*{Code Availability}
The code to perform the analysis will be available upon request.

\bibliography{biblio}

\end{document}


\title{
Supplementary Information for: Mixing Individual and Collective Behaviours to Predict Out-of-Routine Mobility
}

\author{Sebastiano Bontorin}
\affiliation{Fondazione Bruno Kessler, Via Sommarive 18, 38123 Povo (TN), Italy}
\affiliation{Department of Physics, University of Trento, Via Sommarive 14, 38123 Povo (TN), Italy}

\author{Simone Centellegher}
\affiliation{Fondazione Bruno Kessler, Via Sommarive 18, 38123 Povo (TN), Italy}

\author{Riccardo Gallotti}
\affiliation{Fondazione Bruno Kessler, Via Sommarive 18, 38123 Povo (TN), Italy}

\author{Luca Pappalardo}
\affiliation{ISTI - National Research Council, Via Giuseppe Moruzzi 1, 56127 Pisa (PI), Italy}

\author{Bruno Lepri}
\affiliation{Fondazione Bruno Kessler, Via Sommarive 18, 38123 Povo (TN), Italy}

\author{Massimiliano Luca}
\email[Corresponding author:~]{mluca@fbk.eu}%
\affiliation{Fondazione Bruno Kessler, Via Sommarive 18, 38123 Povo (TN), Italy}


\date{\today}

\maketitle

\tableofcontents
\newpage

\section{Dataset Details}

In this Section, we report additional details regarding the dataset used in this study.

\subsection{Details on Cuebiq Dataset for Seattle, Boston and NYC}

The location data is provided by Cuebiq Inc., a location intelligence and measurement company. The dataset was shared within the Cuebiq Data for Good program, which provides access to de-identified and privacy-enhanced mobility data for academic and research purposes. 

The location data provided consists of privacy-enhanced GPS locations for research purposes, from January 2020 to September 2020, and includes only users who have opted-in to share their data anonymously. The data is General Data Protection Regulation (GDPR) and California Consumer Privacy Act (CCPA) compliant. Furthermore, to increase and preserve users' privacy, Cuebiq obfuscates home and work locations to the Census Block Group level. 
The data is collected through the Cuebiq Software Development Kit (SDK) which collects user locations through GPS and Wi-Fi signals in Android and iOS devices.

\begin{table}[htbp]
    \centering
    \begin{tabular}{|l|c|c|}
        \toprule
        city & users & \# points \\
        \midrule
        Seattle & 270K & 12M \\
        Boston & 375K & 11M \\
        NYC & 1,5M & 140M \\
        \bottomrule
    \end{tabular}
    \caption{\small{\textbf{Dataset statistics.} Total number of users and number of spatio-temporal points $p = (i,t)$ in the Cuebiq dataset for the cities of Seattle, Boston and New York City. Numbers have been approximated.}}
\end{table}

        
\subsection{Collective OD flows}

\begin{figure*}[!ht]
    \centering
    \includegraphics[width=1.0\linewidth]{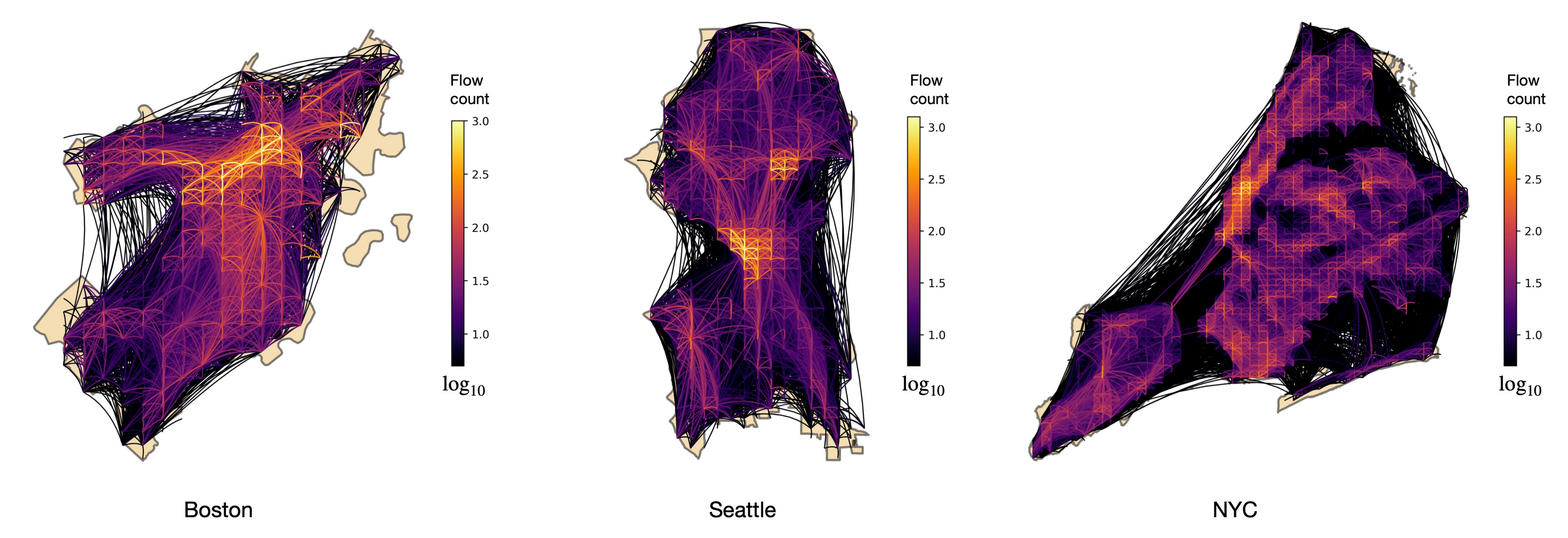}
    \caption{\textbf{OD Flows.} Visualisation of collective flows between Geo Hashes (level 6) in the Cuebiq dataset for Boston, Seattle and NYC in the period from January 3rd to March 1st, 2020.}
\end{figure*}

\clearpage

\subsection{Distribution of Trajectories Overlap}

\begin{figure*}[!ht]
\centering
\includegraphics[width=0.8\linewidth]{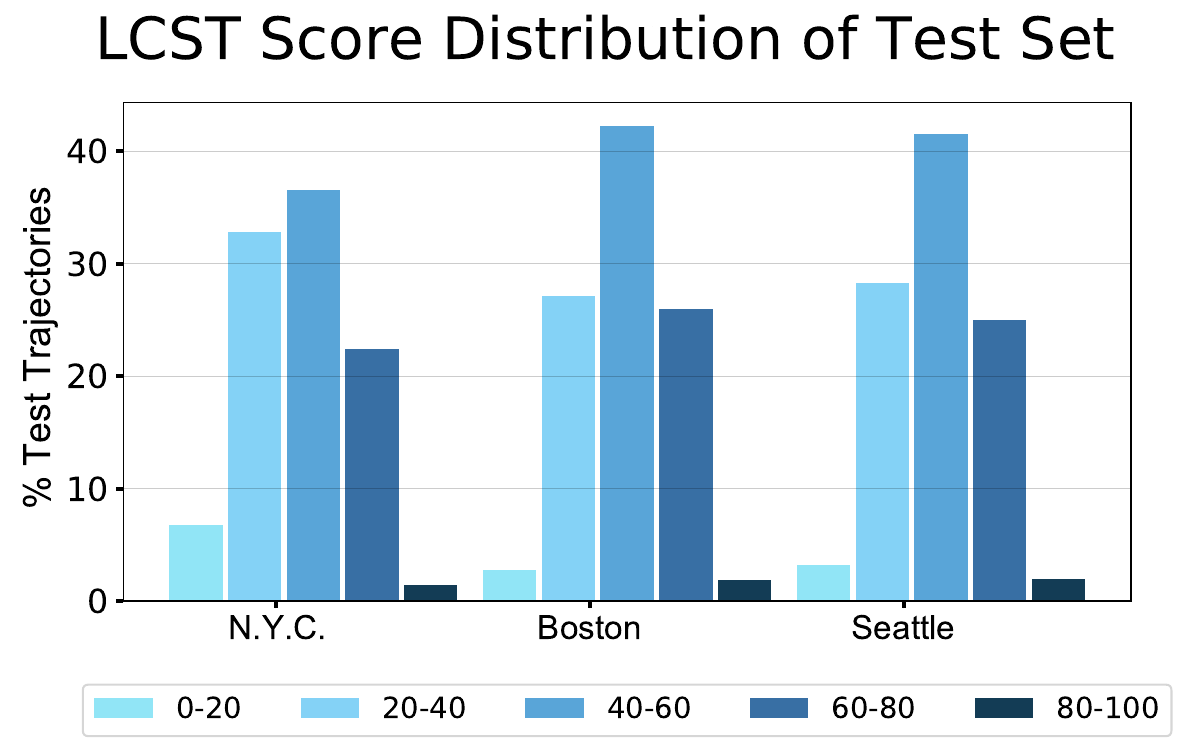}
\caption{\small{\textbf{Distribution of LCST score of the test trajectories for each city.} For each city, we report the percentage of test trajectories that ended up being in one of the five bins we are evaluating. We can observe a similar distribution regardless of the city, with most of the trajectories representing mixed mobility (i.e., 20-80\% overlap with training) while fully novel (i.e. 0-20\% of overlap) and fully observed (i.e., 80-100) are the two less represented mobility profiles.}} 
\label{fig:traj_dist}
\end{figure*}

Percentage of novel transitions never observed during training in a test set are quantified via Longest Common Sub-Trajectory (LCST) score \cite{luca2023trajectory}. LCST scores distribution are displayed in Fig. \ref{fig:traj_dist}.


\subsection{Comparison of 0\% and 0-20\% overlaps.}

\begin{figure*}[!ht]
    \centering
    \includegraphics[width=1.0\linewidth]{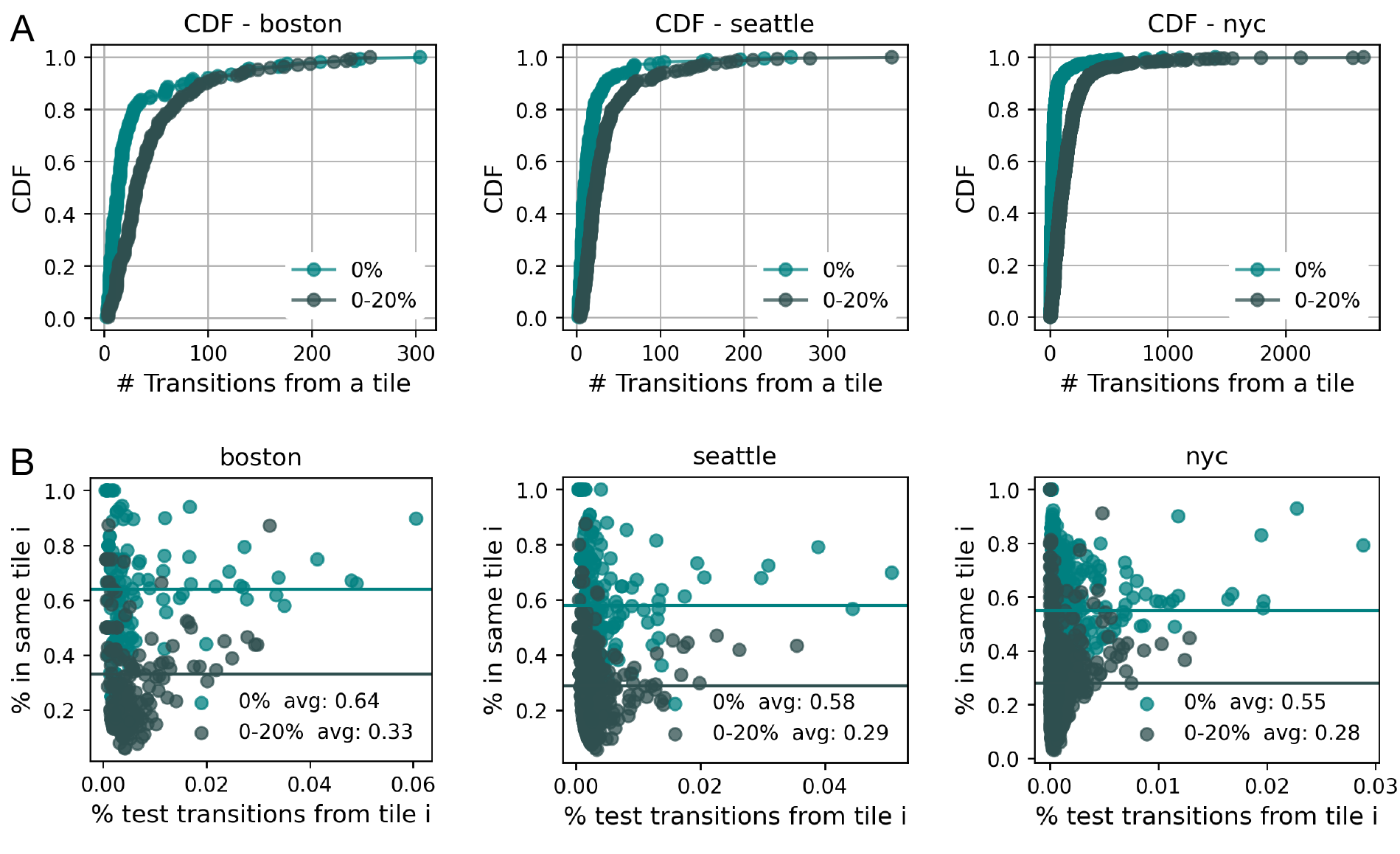}
    \caption{ \textbf{Comparison of overlaps 0\% and 0-20\% test trajectories.}\textbf{(A)} The cumulative distribution function (CDF) of the number of transitions from each location $i$ in the overlaps 0\% and 0-20\%. \textbf{(B)} Scatter plots (each point is a location $i$) of the
    fraction of all test transitions (in 0\% or 0-20\% overlaps) from a location versus the \% of these transitions having both the origin and the destination in the same location $i$ (the individual user remains in the same tile). The weighted average of same-tile transitions across different locations is present in legend, where the weight for each location is the fraction of test transitions having origin in that location.}
    \label{fig:0_overlap}
\end{figure*}

Test transitions in trajectories with an exact 0\% overlap have most of the origin locations concentrated in few tiles (Fig. \ref{fig:0_overlap}A). In Fig. \ref{fig:0_overlap}B we observe that in test overlap 0\%
most transitions have origin and destination in the same location $i$. This indicates that on average the 64\% of the transitions in this overlap are constituted by people that remain in the same location.
We thus decided to exclude trajectories with 0\% overlap as the inclusion of this type of transitions distorts our analysis
and does not allow a fair comparison between the models. We can see in Table \ref{tab:tab_0_acc} models’ ACC@5 on test trajectories having an exact 0\% , as in our dataset are mostly constituted by transitions in few locations which are well predicted by $C$ (and in turn also by $M$), exhibit large accuracy (up to 77\% in Boston).


\begin{table}[h]
\centering
\begin{tabular}{|l|ccc|ccc|ccc|}
\hline
\multirow{2}{*}{City} & \multicolumn{3}{c|}{0\%}   & \multicolumn{3}{c|}{0-20\%} & \multicolumn{3}{c|}{0\% + 0-20\%} \\ \cline{2-10} 
                       & I    & C    & M    & I     & C     & M      & I        & C        & M        \\ \hline
NYC                    & 0.048 & 0.688 & 0.662 & 0.096 & 0.416 & 0.376  & 0.087    & 0.468    & 0.431    \\ 
Boston                 & 0.087 & 0.771 & 0.736 & 0.093 & 0.468 & 0.407  & 0.091    & 0.584    & 0.532    \\ 
Seattle                & 0.089 & 0.664 & 0.635 & 0.073 & 0.395 & 0.34   & 0.078    & 0.48     & 0.433    \\ \hline
\end{tabular}
\caption{\textbf{Accuracies on 0\%, 0-20\% and 0\% + 0-20\% combined.} Models' accuracies on test trajectories with 0\% overlap, with overlap in the 0-20\% bin (0\% excluded) and finally on test trajectories on the full 0-20\% with the 0\% included.}
\label{tab:tab_0_acc}
\end{table}

\clearpage

\subsection{OSM Amenities and POIs retrieval}

OpenStreetMap is a collaborative mapping project that provides rich geographical data, including various amenities types and points of interest. The following list outlines the specific types of amenities extracted from OSM for our analysis:

\begin{center}
   \small{\texttt{\textbf{amenities:['cafe','college','library','university',
'restaurant','pub','fast food','bar','bank','pharmacy','arts centre','cinema','community centre','post office','marketplace']}}} 
\end{center}

The number of POIs extracted from OSM in each GeoHash level 6 tile for each city is presented as a map in Fig. \ref{fig:pr} panel C.

\subsection{User pruning of test trajectories}

We prune over-represented users in the test dataset. We do this by inspection of the distribution of the number of transitions for each user. We consider the $95-th$ percentile (of transitions per user) as a threshold $T^{*}$. 
The pruning consists in the following procedure: if user $u$ has a total number of transitions $T^{(u)}$ in the test set larger than threshold $T^{(u)} > T^{*}$, we randomly remove this user's transitions in the test set until the total number of transitions for any user respects the condition $T^{(u)} \leq T^{*}$.

\clearpage
\newpage

\section{Spatial properties of $I$, $M$ and $C$ accuracy}

\begin{figure*}[!ht]
    \centering
    \includegraphics[width=1.0\linewidth]{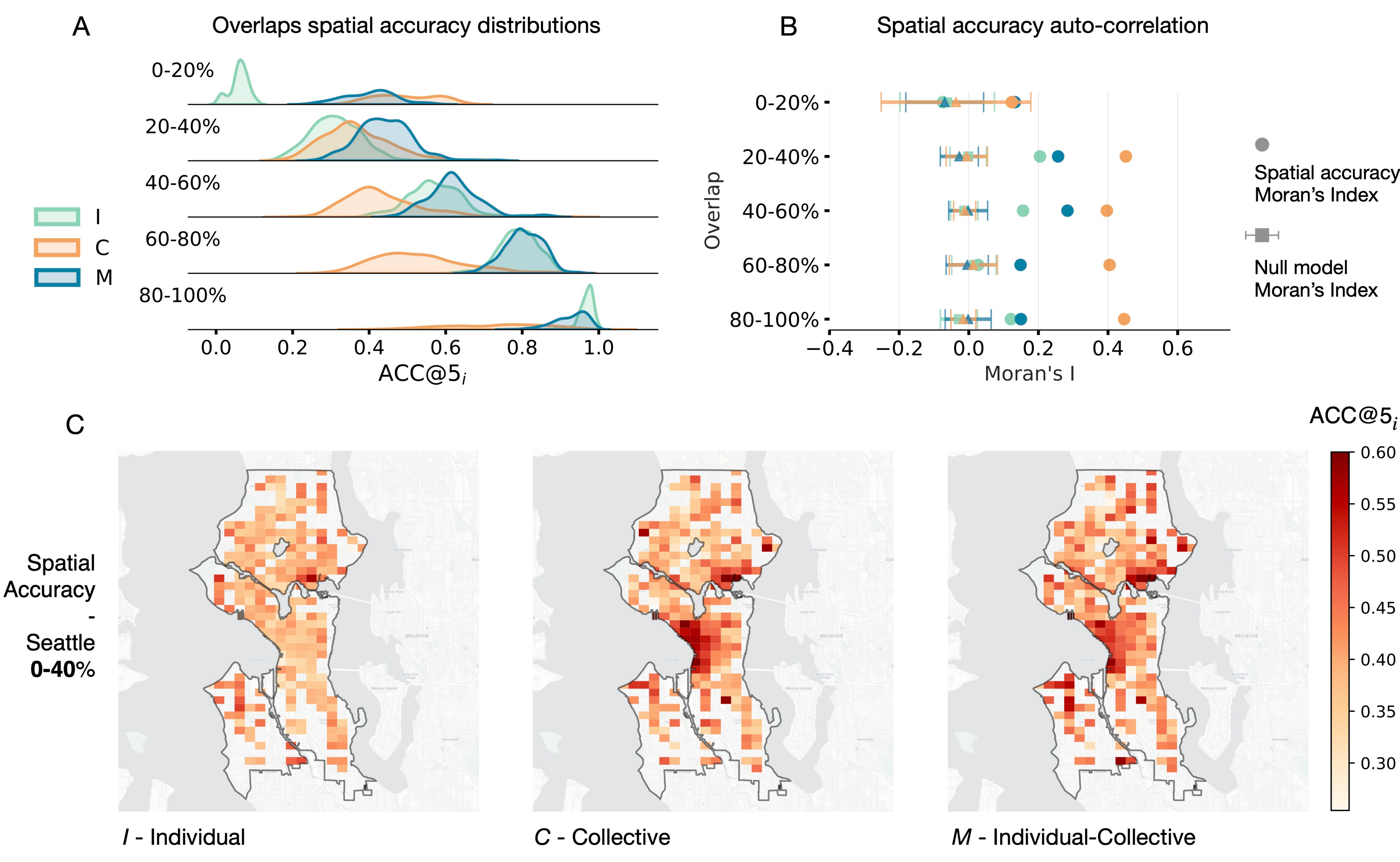}
    \caption{\textbf{Spatial distributions of accuracies - Seattle.} Distribution of accuracies for the individual $I$,the collective $C$ and model $M$ which combines individual and collective information, in predicting a transition from a Geo Hash 6 tile. \textbf{(A)} Spatial accuracies across different overlaps: as the test set includes more novel and out-of-routine mobility (e.g., 0-20\% and 20-40\% overlaps), model $M$ aided by collective information $C$ performs better. While the individual model $I$ provides better predictions when it has seen more trajectories in training, and is tested on recurrent individual patterns. \textbf{(B)} Spatial autocorrelation of the models' accuracies in corresponding overlaps quantified via the Moran Index. For larger out-of-routine behaviours like 0-20\% and 20-40\% overlaps, model $C$ exhibits clustered accuracy (large Moran Index), a property also of model $M$ where collective behaviour improves the predictive capabilities. \textbf{(C)} Map of spatial accuracies ACC@5$_{i}$ in the overlap 0-40\% in Seattle for models $I$,$C$ and $M$. Maps: Stamen Maps}
    \label{fig:spatial_acc}
\end{figure*}

\begin{figure*}[!ht]
    \centering
    \includegraphics[width=1.0\linewidth]{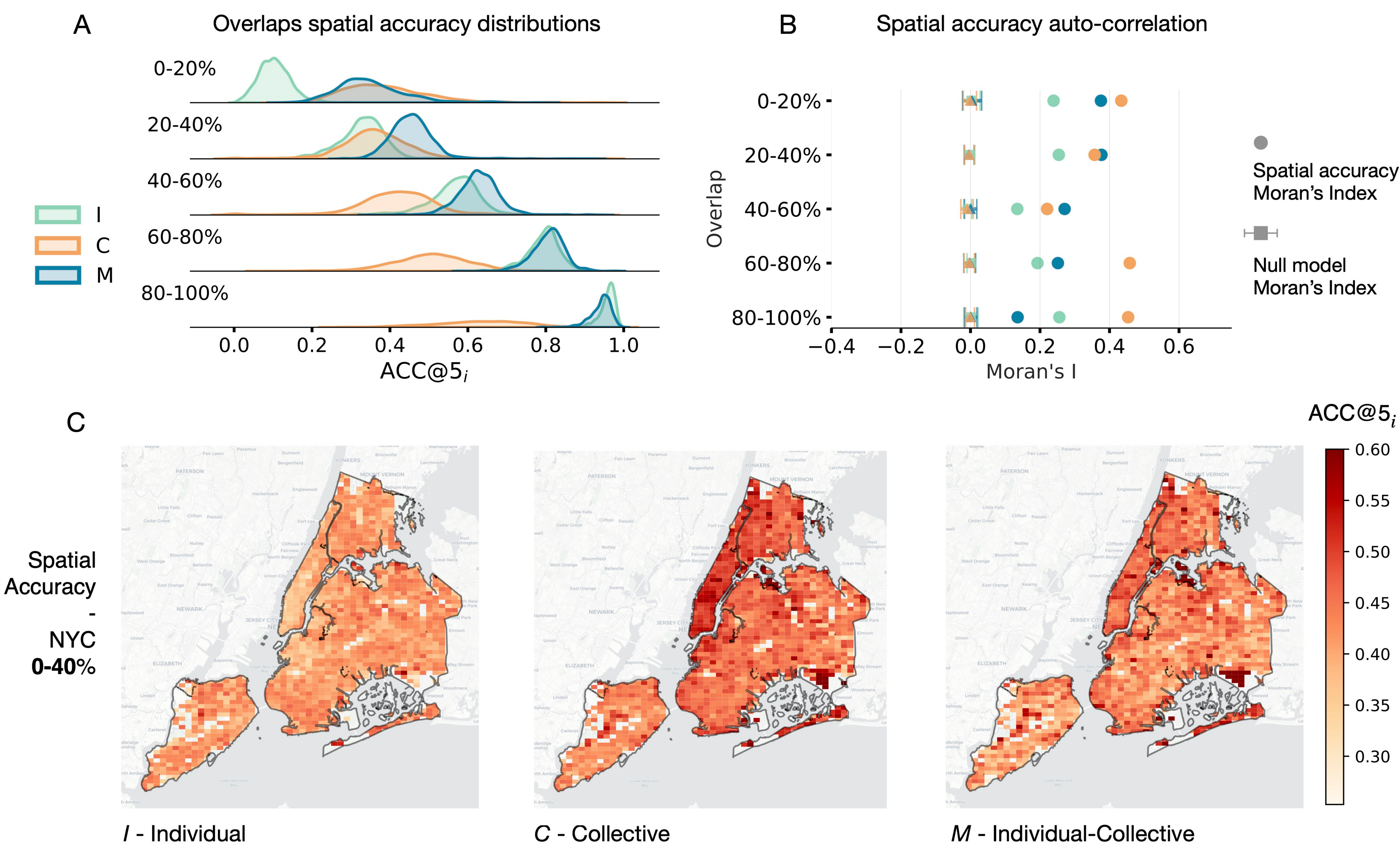}
    \caption{\textbf{Spatial distributions of accuracies - NYC.} Distribution of accuracies for the individual $I$,the collective $C$ and model $M$ which combines individual and collective information, in predicting a transition from a Geo Hash 6 tile. \textbf{(A)} Spatial accuracies across different overlaps: as the test set includes more novel and out-of-routine mobility (e.g., 0-20\% and 20-40\% overlaps), model $M$ aided by collective information $C$ performs better. While the individual model $I$ provides better predictions when it has seen more trajectories in training, and is tested on recurrent individual patterns. \textbf{(B)} Spatial autocorrelation of the models' accuracies in corresponding overlaps quantified via the Moran Index. For larger out-of-routine behaviours like 0-20\% and 20-40\% overlaps, model $C$ exhibits clustered accuracy (large Moran Index), a property also of model $M$ where collective behaviour improves the predictive capabilities. \textbf{(C)} Map of spatial accuracies ACC@5$_{i}$ in the overlap 0-40\% in NYC for models $I$,$C$ and $M$. Maps: Stamen Maps}
    \label{fig:spatial_acc}
\end{figure*}

\clearpage
\newpage


\section{Statistical properties of Collective mobility $C$ in proximity to POIs}

\begin{figure*}[ht!]
\centering
\includegraphics[width=0.9\linewidth]{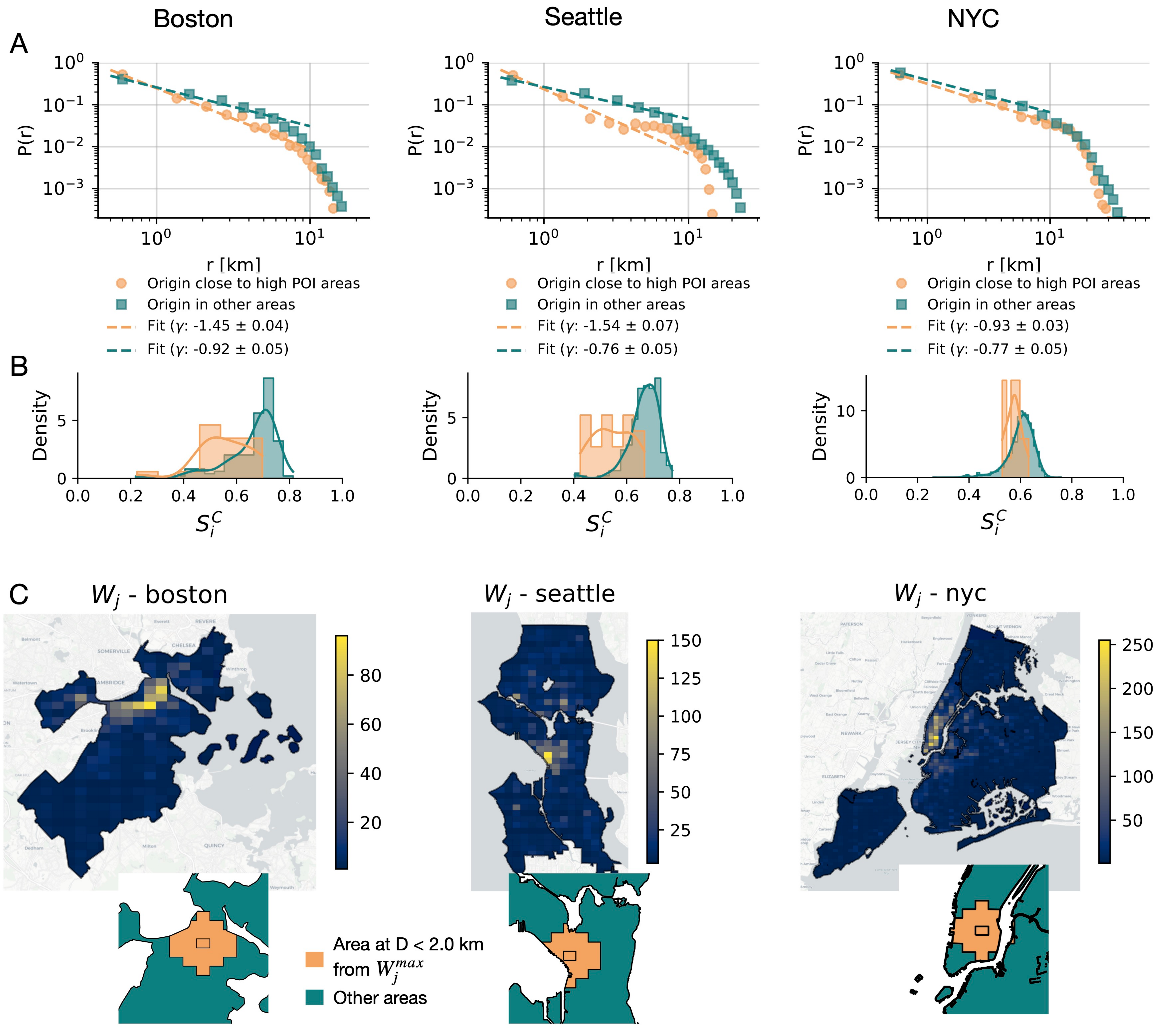}
\caption{\small{\textbf{Statistical properties of Collective mobility and proximity of Points of Interests.}
    We separate the cities into tiles within a distance D from the attractiveness centre (location with largest number of POIs $W_{j}^{max}$) and other areas with distance greater than D.
    \textbf{(A)} the distribution of travel distances ($P(r)$), distinguishing between origins within D = 2 km from the area with highest number of POIs ($W_{j}^{max}$) and other origins. Distribution are fitted with a power-law function $P(r) \sim r^{-\gamma}$ in the interval of 0 to 10 Km. The different exponents $\gamma$ highlight a more localised mobility in proximity of POIs.
    \textbf{B)} Entropy $S_{i}^{C}$ distributions for locations in the two areas.
    \textbf{C)} Map of the number of POIs ($W_{i}$) in a Geo hash 6 tile $i$, extracted from OpenStreetMap.
    \textbf{(Inset)} Areas within a distance $D = 2$ Km from $W_{j}^{max}$ are shaded in orange. In teal other areas.
    Albeit this separation of the city is solely based on proximity to high POIs areas, it separates collective human mobility in two different predictability regimes. Maps: Stamen Maps}}
\label{fig:pr}
\end{figure*}

\begin{figure*}[h!]
\centering
\includegraphics[width=1.\linewidth]{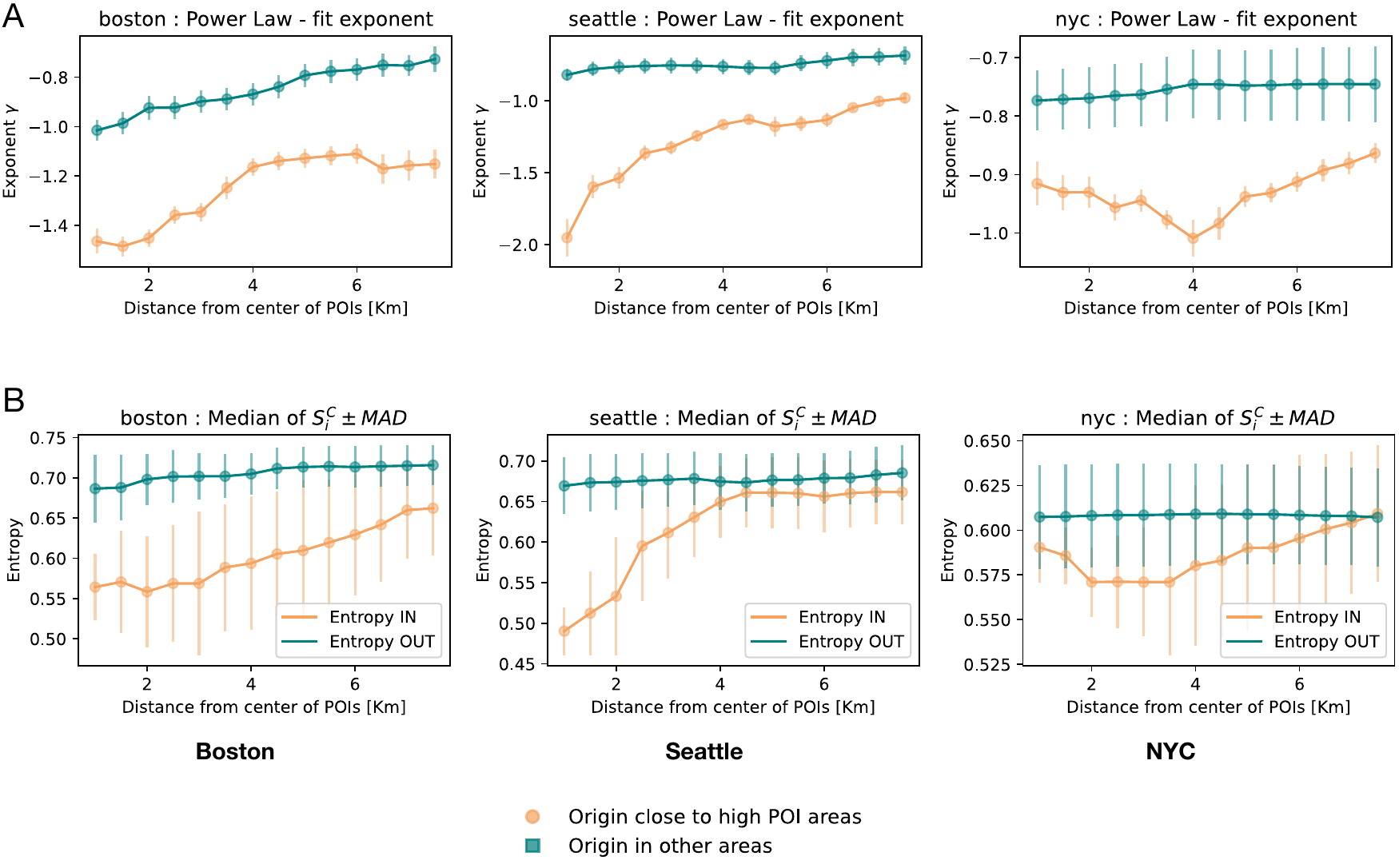}
\caption{\small{\textbf{Sensitivity analysis of travel distances $P(r)$ and collective $S_{i}^{C}$ in proximity of POIs and in other areas.} \textbf{A)} Exponents $\gamma$ of power law fits (in the range 0 to 10 km) of the travel distributions $P(r)$. In orange the exponents for $P(r)$ computed on transitions having location origin within distance D from the tile $W_{i}^{max}$, and $P(r)$ for origin in other areas. \textbf{B)} The median value and its median absolute deviation (MAD) of entropy $S_{i}^{C}$ distributions for locations within D from $W_{i}^{max}$ or in other areas.
For Seattle and NYC in particular we observe a range of threshold distances D in which we can appreciate a separation of exponents and median values. In particular D = 2 Km represents a proximity distance where in all cities we observe different statistical properties.}}
\label{fig:flows_direction}
\end{figure*}

\begin{figure*}[h!]
\centering
\includegraphics[width=1.\linewidth]{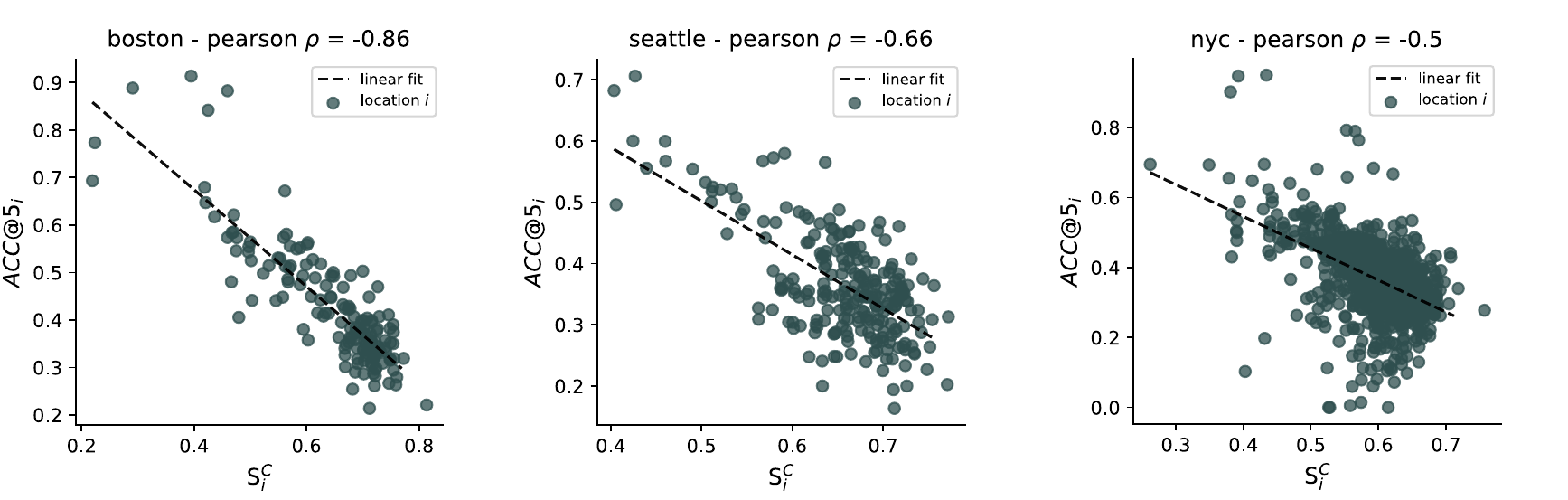}
\caption{\small{\textbf{Pearson correlation of $S_{i}^{C}$ versus ACC@5$_{i}$.} We compute the Pearson correlation $\rho$ of accuracy of $C$ model from a tile $i$ ACC@5$_{i}$ in the 0-40\% overlap versus the collective entropy $S_{i}^{C}$ in that location. We observe negative correlation between the two variables, specifically for Boston $\rho = -0.86$, Seattle $\rho = -0.66$ and NYC $\rho = -0.5$. }}
\label{fig:pearson_entropy_acc}
\end{figure*}

\begin{figure*}[!]
\centering
\includegraphics[width=1.0\linewidth]{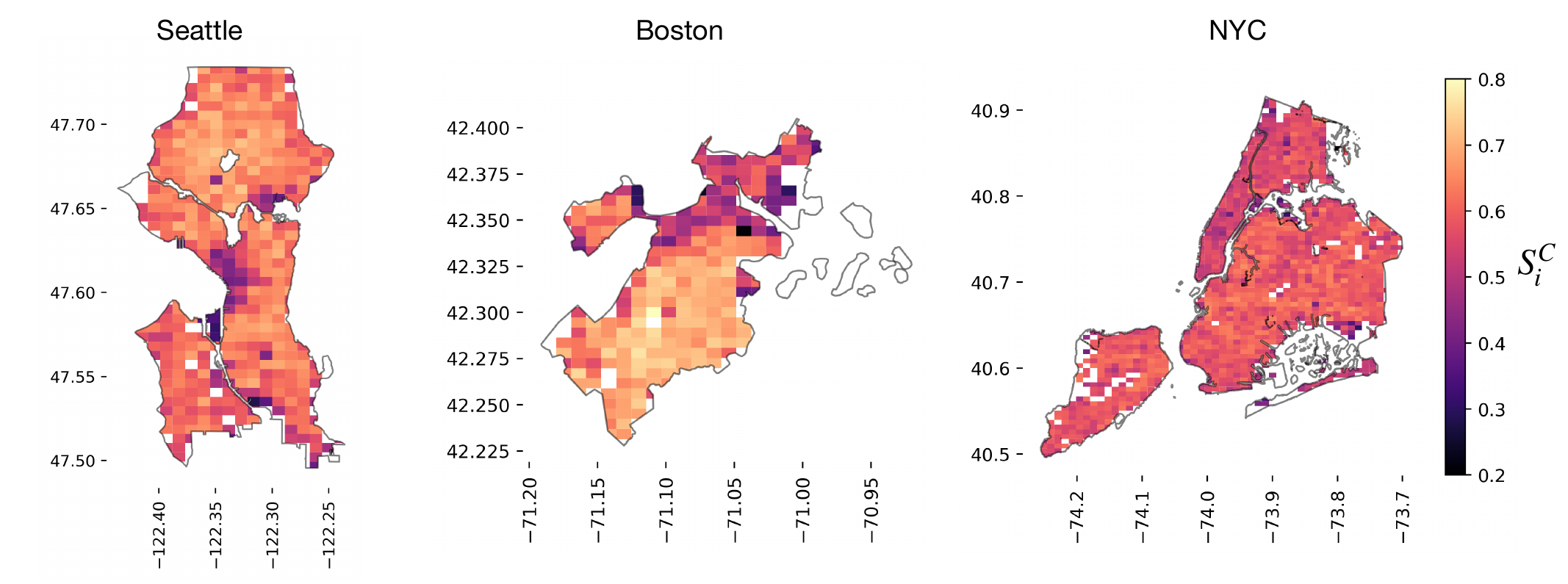}
\caption{\small{\textbf{Collective Entropy $S_{i}^{C}$ at Geo Hash level 6.} Spatial distribution of normalised entropy $S_{i}^{C}$ of the Collective OD. Areas in violet are characterised by lower entropy, with destination probabilities from location $i$ being concentrated in a smaller set of tiles. This also results in a larger predictability and a stronger improvement of predictions of out-of-routine mobility.} }
\label{fig:collective_entropy}
\end{figure*}

\clearpage

\section{Robustness of the predictive capabilities of collective behaviours.}

Areas in Fig. \ref{fig:pr} with high density of POIs and in which novel mobility is better predicted, are also areas characterised by larger number of transitions in the dataset.
We conducted a robustness test to verify that the accuracy improvements in these urban areas are not attributable to probabilities $C$ which may benefit from a larger sample size.
Specifically, we reduce and control the collective information that can be accessed by our model $M$ by pruning the set of transitions from these over-represented locations used to estimate the collective OD. In the following, we describe in detail this process.

\begin{figure*}[h!]
\centering
\includegraphics[width=1.0\linewidth]{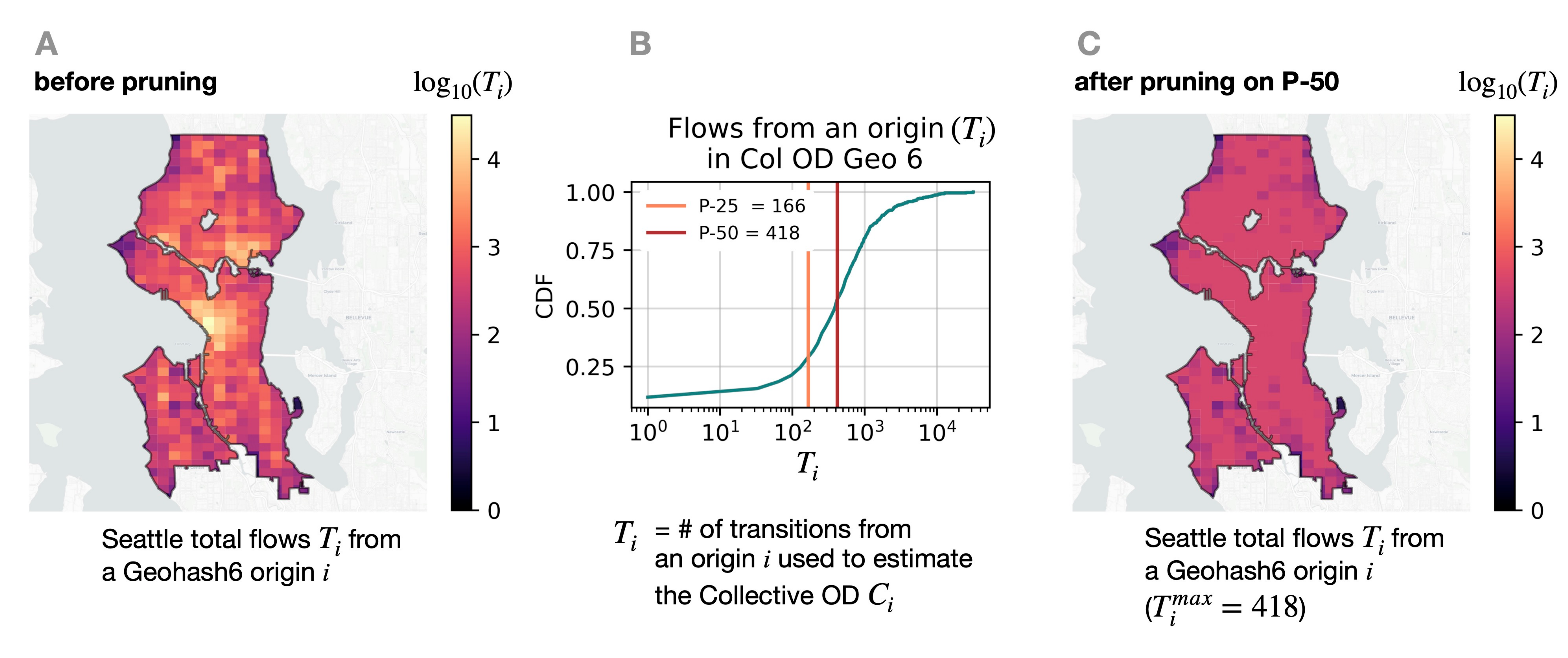}
\caption{\small{\textbf{Sub-sampling of Collective OD flows and reduction of sample-size effects.} A) the number for total transitions in the training dataset from a Geo hash tile $i$ for Seattle, computed as $T_{i} = \sum_{j}T_{ij}$ where $T_{ij}$ is the flow count of transitions from location $i$ to location $j$. B) The cumulative distribution function of $T_{i}$ highlights that several orders of magnitude separate few locations from the remaining areas. The 25-th and 50-th percentiles are highlighted. C) After the pruning process in Algorithm \ref{algo:algo1} the value $T_{i}$ in all tiles is approximately similar. (Here $T_{i}^{max}$ = 418)}}
\label{fig:pruning}
\end{figure*}

\subsection{Pruning of over-represented users in location $i$ in $C$ estimation.}

First, we remove biases introduced by over-represented users in the OD. If a user $u$ has a number of transitions from an origin location $i$ larger than the $50-th$ percentile of users in location $i$ (which we define as $T^{u-50}_{i}$), we prune its transitions in $i$ up to the threshold $T^{u-50}_{i}$. After pruning, the number of transitions for each user in location $i$ is therefore $T^{(u)}_{i} \leq T^{u-50}$.

\subsection{Pruning of Collective OD in over-represented locations $i$.}

Second, we focus on the total number of transitions in the training set from location $i$ available to reconstruct probabilities $C_{i}$.
If the number of training transitions from location $i$ exceeds the median ($50-th$ percentile) of all locations (which we refer to as $T^{max}$), we uniformly and randomly pruned the subset of training transitions from location $i$ until this threshold was reached ($T_{i} < T^{max}$). This process allows to perform next location predictions with collective $C_{i}$ where the probabilities have been computed with a comparable amount of information across all tiles $i$.
The process of pruning origins is presented in Algorithm \ref{algo:algo1}.

\begin{algorithm}
    \BlankLine
    \SetAlgoLined
    \KwData{Set of training transitions $D_{i}$ for collective OD estimation from origin $i$}
    \KwData{Size of the training set from location $i$: $T_{i} = |D_{i}| = \sum_{j} T_{ij}$}
    \KwData{Percentile $X$ (in the analysis $X=50\%$)} 
    \KwResult{Pruned collective OD $\bar{C_{i}}$}

    \BlankLine
    \textbf{Compute CDF and determine percentile $X$ transitions $P_{X}$}\;
    
    \ForEach{origin $i$}{
        \If{$T_{i}$ is larger than $X$ percentile transitions $P_{X}$ ($T^{max}$)}{
            \textbf{Sample uniformly} $P_{X}$ transitions from $D_{i}$, as sub-sampled set $\bar{D}_{i}$\;
            
            \textbf{Estimate} new pruned collective probabilities $\bar{C_{i}}$ using sub sample $\bar{D}_{i}$\;
        }
    }
    \BlankLine
    \BlankLine
    \caption{\small{\textbf{Pruning Collective OD Estimation.} Pruning algorithm to remove bias of collective OD $C_{i}$ being estimated with a larger sample size in locations close to high density of POIs.}}
    \label{algo:algo1}
\end{algorithm}

\begin{figure*}[h!]
\centering
\includegraphics[width=1.0\linewidth]{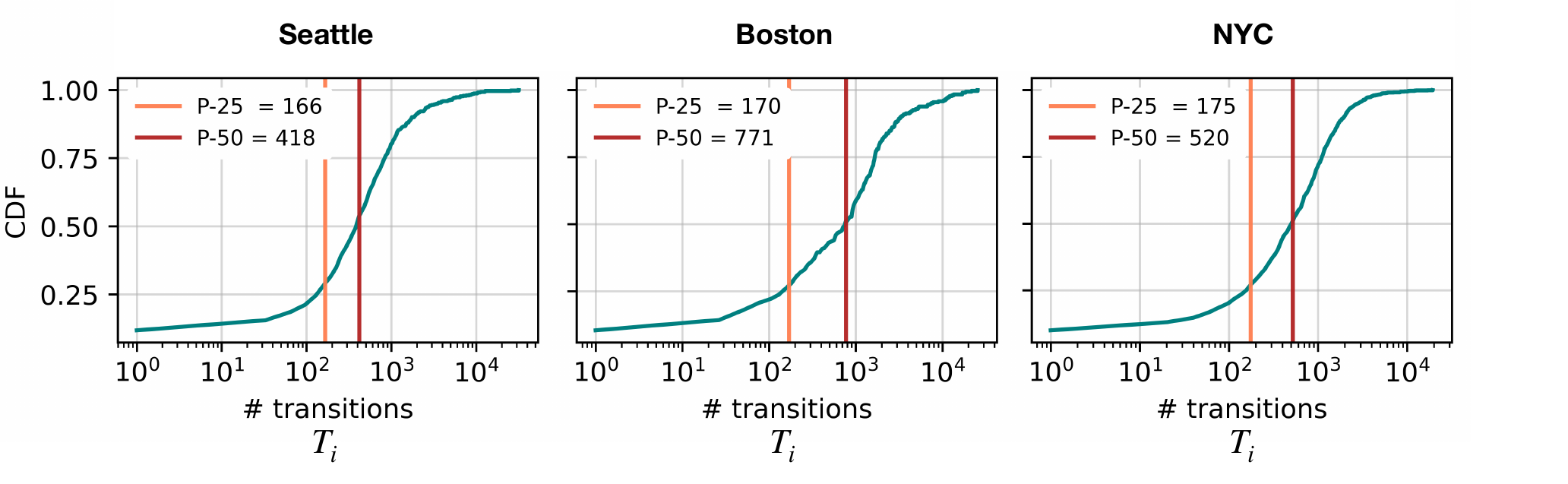}
\caption{\small{\textbf{Cumulative distribution of transition per origin in the train dataset.} 25-th and 50-th percentile CDF values are indicated.}}
\label{fig:CDF}
\end{figure*}

Moreover in Fig. \ref{fig:pruning} we present the effect of pruning over-represented origins in the training dataset. We show the case of Seattle, where the number of available flows $T_{i}$ from Geo Hashes $i$ spans different orders of magnitude for different areas in the original dataset. After computing the CDF and percentiles (panel B) we prune the set of training transitions following algorithm \ref{algo:algo1} based on the median value as threshold. The resulting set of comparable flows $T_{i}$ across tiles after a pruning procedure is in panel C.
All CDF and percentiles used for the set of three cities are reported in Fig. \ref{fig:CDF}.

\begin{figure*}[h!]
\centering
\includegraphics[width=1.0\linewidth]{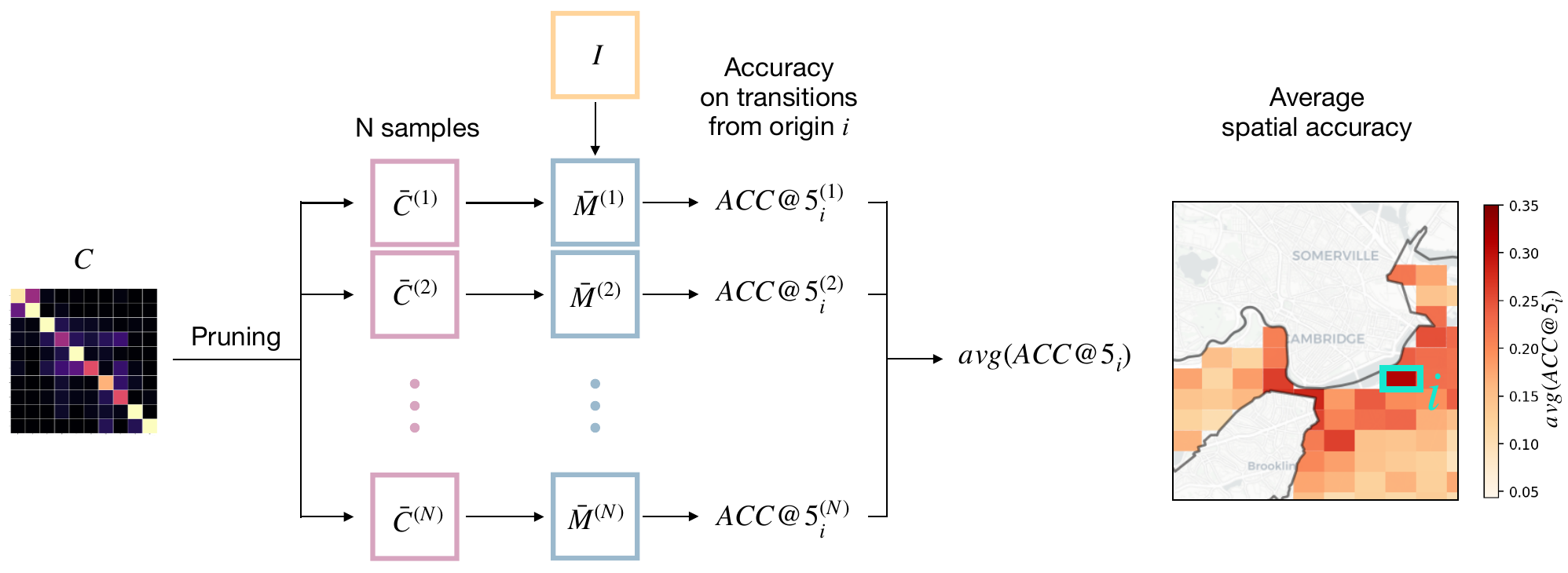}
\caption{\small{\textbf{Sub-sampled $\bar{C}^{(i)}$ and computation of average spatial accuracy from pruned ODs.} Procedure of generation of N samples of pruned collective ODs and $\bar{C}^{(i)}$, and computation of average value of accuracy from an origin location $i$ (ACC@5$_{i}$) across the sub-sampled models.}}
\label{fig:sampling_procedure}
\end{figure*}

\subsection{Spatial accuracies for Origin and Destination}

This random sub-sampling process was repeated for a number of samples $N_{samples} =10$, generating sub-datasets from which $N_{samples}$ collective Markov models $C_{i}$ have been computed. From the sub-sampled $\bar{C}$, different models $M$ have been tested on the dataset and a final average on spatial accuracy in each Geo Hash 6 tile has been obtained. We indicate sub sample $i$ model as $\bar{C}^{(i)}$. This sampling procedure and computation of average spatial accuracy is reported in Fig. \ref{fig:sampling_procedure}

\begin{figure*}[h!]
\centering
\includegraphics[width=1.0\linewidth]{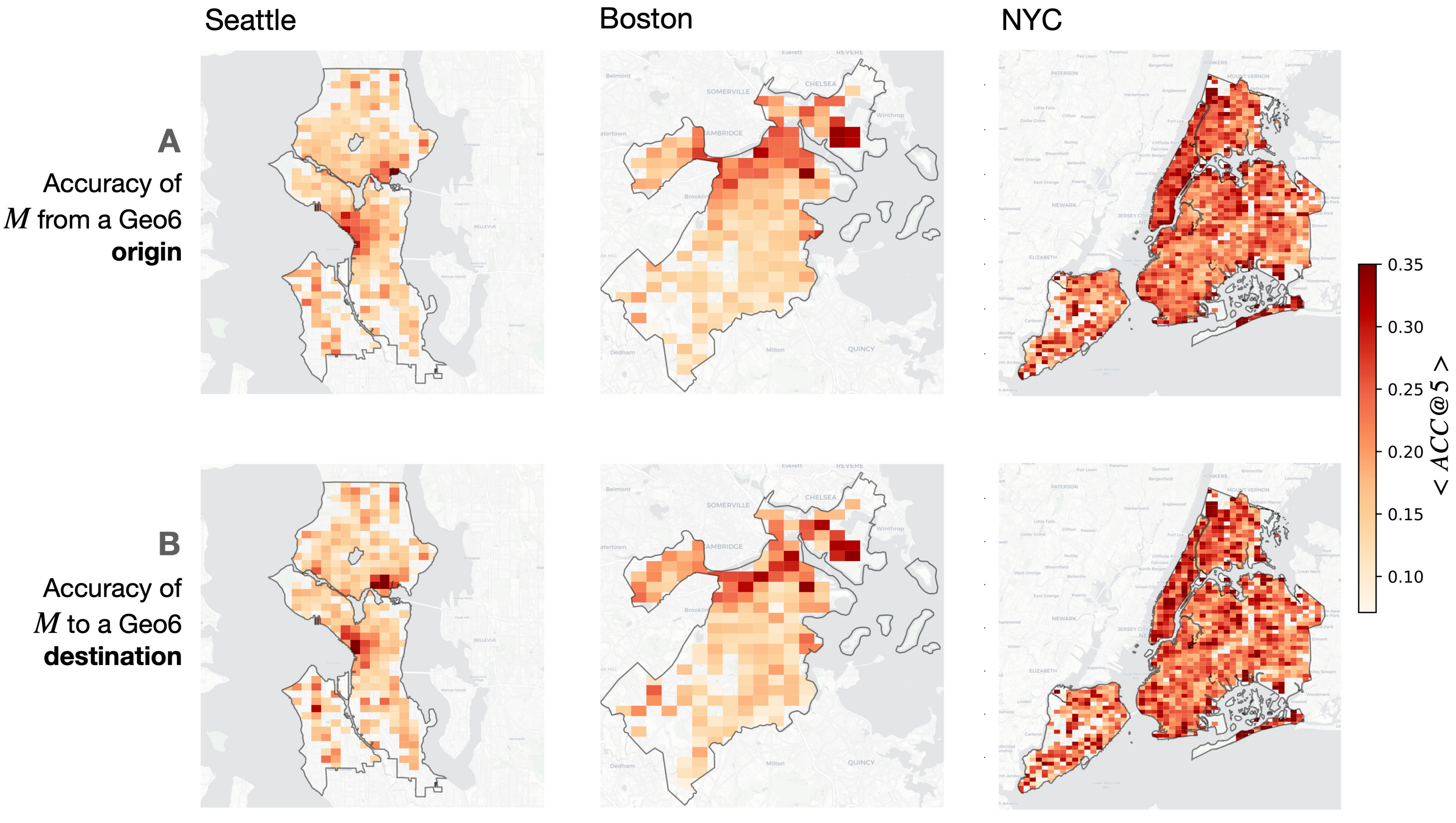}
\caption{\small{\textbf{Spatial accuracy of $M$ model on novel mobility with pruned $\bar{C}$ in Seattle, Boston and NYC.} The accuracy of $M$ model in predicting a novel transition (never seen in a individual $u$'s historical trajectories $\mathcal{H}^{(u)}$) from an origin in a Geo Hash $i$ is presented in panel $\textbf{(A)}$. In panel $\textbf{(B)}$ the accuracy in predicting a location as destination of a test transition is shown. It's worth noting that the sample size used to estimate $C_{i}$ exhibits strong differences, with central areas being more densely sampled. To mitigate this potential bias, we employed a stochastic sub sampling process to estimate $C_{i}$. Thus, spatial heterogeneity in ACC@5$_{i}$ cannot be attributed to differences in the density and richness of the dataset used for estimating $C_{i}$ probabilities. Maps: Stamen Maps}}
\label{fig:ACC}
\end{figure*}

\begin{figure*}[h!]
\centering
\includegraphics[width=1.0\linewidth]{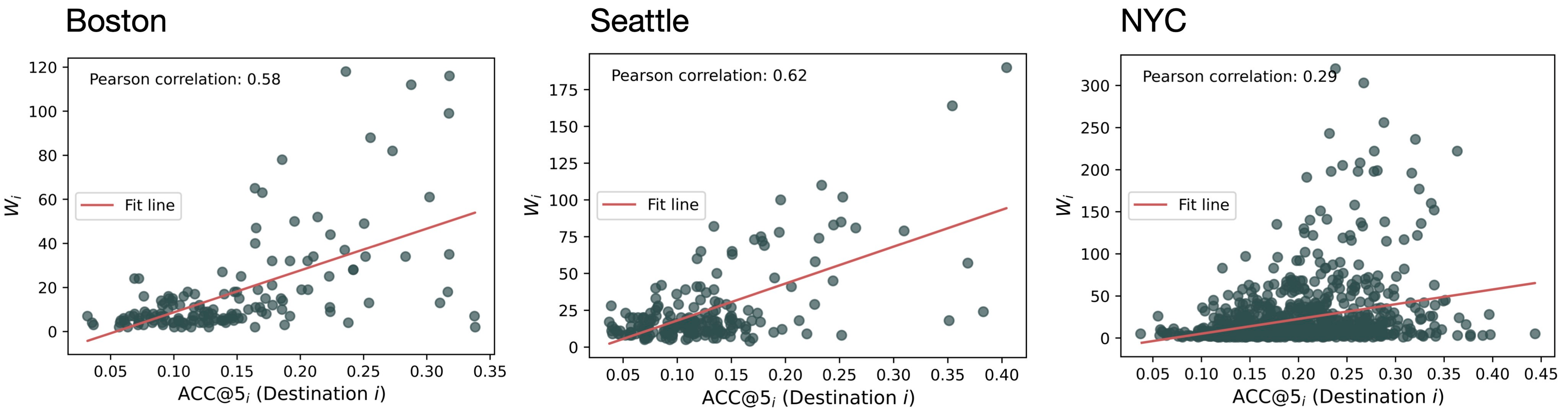}
\caption{\small{\textbf{Pearson coefficient between $ACC@5_{i}$ (Destination $i$) and $W_{i}$.} Correlation between accuracy in predicting a destination location and the number of POIs in that location for the three cities. We observe that, specifically for Boston and Seattle, the best predicted destinations in the regime of novel mobility (overlap 0-40\%) are the locations having the largest number of POIs. A lower Pearson is observed for New York City, where also locations with lower density of POIs are well predicted as destinations. }}
\label{fig:ACC_pearson}
\end{figure*}

In Fig. \ref{fig:ACC}, we report the spatial accuracies of $M$ using pruned $\bar{C}$. Here we adopt exclusively novel transitions extracted from the overlap $0-40\%$, i.e. instances in which the correct destination $j$ from a location $i$ in the test has never been seen in $I_{i}^{(u)}$ for user $u$. This is done to investigate the spatial dependencies of pruned collective information $\bar{C}$ to aid the $M$ model when individual patterns can not aid by definition because they lack the necessary information. Therefore, eventual heterogeneity in accuracy observed can be attributable exclusively to collective information.
We present accuracies both from an origin location $i$ (panel A) and the accuracy in estimating correctly a destination to a tile $j$ (panel B) using the model $M$ built from Collective OD pruned $\bar{C}$, as described in Algorithm \ref{algo:algo1} and Fig. \ref{fig:pruning}.\\

Spatial accuracies presented in Fig. \ref{fig:ACC} are obtained as the mean value of ACC@5$_{i}$ across the ensemble of sub sampled models for each location. We observe that the enhanced ACC@5$_{i}$ accuracy in of $M$ and $C$ in proximity to areas with high density of POIs is still present even after pruning the information with which $C_{i}$ is computed. Therefore the improvement in accuracy can not be attributed to the number of transitions  in the training dataset used to estimate $C_{i}$ probabilities. Moreover we observe that the best predicted destinations $j$, with high ACC@5$_{j}$, seem to also be clustered around high POIs areas. Therefore we compute the Pearson correlation between the accuracy ACC@5$_{j}$ of model $M$ in predicting correctly a a location $j$ as a destination and the number of POIs $W_{j}$ in that destination. In Fig. \ref{fig:ACC_pearson} we show these scatter plots for Seattle, Boston and NYC. Areas with large number of POIs tend to be the ones also better predicted as destinations by the collective behaviours. This trend in particular is more pronounced for Boston and Seattle, with a $\rho = 0.58$ and $\rho = 0.62$ respectively.




\clearpage

\newpage

\section{Recurrent Neural Network Implementation}
\label{sec:si_rnn}
RNNs are commonly used as baselines for tasks in which sequential information is involved. Given a sequence as input, an RNN performs the same task for each element and the output depends on the previous computation. Each computation involves three parameters:
\begin{itemize}
    \item $x_i$ the input at the $i^{th}$ step
    \item $h_i$ the hidden layer at the $i^{th}$ step
    \item $y_i$ the output of the $i^{th}$ step
\end{itemize}
There are many different types of recurrent neural networks and for this work, we leveraged an Elman RNN as implemented in \cite{feng2018deepmove}. The computation performed by the networks at with $n$ gates are the following: 
$$ h_i = \sigma{_h} (W_{h_i} + U_{h_i} h_{i-1} + b_{h_i}), \text{for each } i \in \{1, \dots, n-1\} $$
$$ y_i = \sigma{_y} (W_{h_{n-1}} + b_{n-1}) $$
To train the RNN, we fine-tuned the following hyperparameters:  learning rate (0.001 for New York, 0.005 for Seattle and Boston),  hidden size (750), embedding size (400), and epochs (250 with early stop mechanism). We use Adam as an optimizer.


\section{Accuracies Tables}

\begin{table*}[htbp]
    \centering
    \resizebox{\textwidth}{!}{
        \begin{tabular}{|l|l|llll|llll|llll|llll|llll|llll|llll}
            \hline
            \textbf{}              & \textbf{} & \multicolumn{4}{c|}{Full Set} & \multicolumn{4}{c|}{0-20}     & \multicolumn{4}{c|}{20-40}    & \multicolumn{4}{c|}{40-60}    & \multicolumn{4}{c|}{60-80}    & \multicolumn{4}{c|}{80-100}                         \\ \hline
            &           & $I$   & $C$   & $M$   & RNN   & $I$   & $C$   & $M$   & RNN   & $I$   & $C$   & $M$   & RNN   & $I$   & $C$   & $M$   & RNN   & $I$   & $C$   & $M$   & RNN   & $I$   & $C$   & $M$   & \multicolumn{1}{l|}{RNN}   \\ \hline
            \multirow{3}{*}{Geo 6} & NYC       & 0.608 & 0.503 & 0.678 & 0.649 & 0.096 & 0.416 & 0.376 & 0.169 & 0.319 & 0.398 & 0.468 & 0.328 & 0.572 & 0.453 & 0.637 & 0.599 & 0.801 & 0.546 & 0.817 & 0.925 & 0.966 & 0.698 & 0.948 & 0.979 \\
                                       & Boston    & 0.706 & 0.643 & 0.753 & 0.746 & 0.093 & 0.468 & 0.407 & 0.155 & 0.32 & 0.454 & 0.492 & 0.366 & 0.604 & 0.537 & 0.68 & 0.637 & 0.831 & 0.669 & 0.847 & 0.928 & 0.977 & 0.839 & 0.929 & 0.98 \\
                                       & Seattle   & 0.645 & 0.549 & 0.697 & 0.694 & 0.073 & 0.395 & 0.34 & 0.148 & 0.314 & 0.397 & 0.461 & 0.343 & 0.581 & 0.471 & 0.646 & 0.625 & 0.808 & 0.584 & 0.824 & 0.936 & 0.971 & 0.781 & 0.918 & 0.969 \\ \hline
            \multirow{3}{*}{Geo 7} & NYC       & 0.475 & 0.329 & 0.530 & 0.494 & 0.063 & 0.221 & 0.222 & 0.108 & 0.244 & 0.229 & 0.333 & 0.26 & 0.508 & 0.305 & 0.547 & 0.529 & 0.757 & 0.424 & 0.768 & 0.77 & 0.951 & 0.616 & 0.939 & 0.969 \\
                                       & Boston    & 0.562 & 0.446 & 0.607 & 0.582 & 0.07 & 0.264 & 0.254 & 0.103 & 0.259 & 0.285 & 0.364 & 0.277 & 0.551 & 0.394 & 0.594 & 0.583 & 0.787 & 0.534 & 0.794 & 0.801 & 0.962 & 0.745 & 0.919 & 0.97 \\
                                       & Seattle   & 0.530 & 0.412 & 0.571 & 0.548 & 0.05 & 0.223 & 0.214 & 0.099 & 0.255 & 0.262 & 0.344 & 0.275 & 0.524 & 0.368 & 0.56 & 0.54 & 0.764 & 0.504 & 0.77 & 0.779 & 0.958 & 0.74 & 0.916 & 0.966 \\ \hline
        \end{tabular}}
    \caption{Full accuracies table. ACC@5 results are presented for both Geo Hash 6 and Geo Hash 7 tessellations.}
    \label{tab:yourlabel}
\end{table*}

\clearpage

\bibliography{biblio}
